\documentclass[aps,prc,preprint,nofootinbib,floatfix]{revtex4-1}
\usepackage{graphicx,xspace,multirow,amsmath,amssymb}
\usepackage[colorlinks=true,citecolor=blue,linkcolor=blue,urlcolor=blue,filecolor=blue]{hyperref}

\newcommand{\pt}{\mbox{$p_{\rm T}$}\xspace}

\newcommand \sqsn{\mbox{$\sqrt{s_{_{NN}}}$}\xspace}

\def\mean#1{\ensuremath{\left<#1\right>}}

\newcommand{\sqrtsnn}{\ensuremath{\sqrt{s_{_{NN}}}}\xspace}
\newcommand{\UA}{\ensuremath{U_{A}(1)}\xspace}
\newcommand{\lambdas}{\ensuremath{\lambda_{*}}\xspace}
\newcommand{\lambdasmax}{\ensuremath{\lambda_{*}^{max}}\xspace}
\newcommand{\lamfrac}{\ensuremath{\lambdas(\mT)/\lambdasmax}\xspace}
\newcommand{\mT}{\ensuremath{m_\mathrm{T}}\xspace}
\newcommand{\PT}{\ensuremath{p_\mathrm{T}}\xspace}
\newcommand{\etap}{\ensuremath{{\eta^\prime}}\xspace}
\newcommand{\metap}{\ensuremath{m_{\etap}}\xspace}
\newcommand{\meps}{\ensuremath{m_{\etap}^{*}}\xspace}
\newcommand{\Tfo}{\ensuremath{T_{FO}}\xspace}
\newcommand{\Teff}{\ensuremath{T_{eff}}\xspace}
\newcommand{\Tcond}{\ensuremath{T_{cond}}\xspace}
\newcommand{\Binv}{\ensuremath{B^{-1}}\xspace}
\newcommand{\uT}{\ensuremath{\mean{u_\mathrm{T}}}\xspace}
\newcommand{\GeV}{\ensuremath{\mathrm{GeV}}\xspace}
\newcommand{\MeV}{\ensuremath{\mathrm{MeV}}\xspace}
\newcommand{\pip}{\ensuremath{\pi^+}\xspace}
\newcommand{\pim}{\ensuremath{\pi^-}\xspace}
\newcommand{\ks}{\ensuremath{\mathrm{K}^{0}_{S}}\xspace}
\newcommand{\omepip}{\ensuremath{\omega\rightarrow\pip}\xspace}
\newcommand{\etapip}{\ensuremath{\eta\rightarrow\pip}\xspace}
\newcommand{\etappip}{\ensuremath{\etap\rightarrow\pip}\xspace}
\newcommand{\kspip}{\ensuremath{\ks\rightarrow\pip}\xspace}
\def\p#1{\ensuremath{{{\bf p}_{#1}}}\xspace}
\def\NmT#1#2{\ensuremath{N^{#1}_{#2}(\mT)}\xspace}

\begin{document}


\title{Significant in-medium $\etap$ mass reduction in $\sqrt{s_{NN}}=200$ GeV Au+Au collisions                                 at the BNL Relativistic Heavy Ion Collider}

\author{R. V\'ertesi$^1$, T.~Cs\"org\H{o}$^{2,1}$ and J.~Sziklai$^1$}
\affiliation{$^1$MTA KFKI RMKI, H--1515 Budapest 114, P.O.~Box~49, Hungary \\ 
$^2$Department of Physics, Harvard University, 17 Oxford St, Cambridge, Massachussetts 02138, USA}

\date{\today}

\begin{abstract}
In high energy heavy ion collisions a hot and dense medium is formed, where the $\mathrm{U_A}(1)$ 
or chiral symmetry may temporarily be restored. As a consequence, the mass of the $\etap$(958) mesons 
may be reduced to its quark model value, and the abundance of $\etap$ mesons at low $p_T$ may
be enhanced by more than a factor of 10. The intercept parameter $\lambdas$ of the charged pion 
Bose--Einstein correlations provides a sensitive observable of the possibly enhanced $\etap$ abundance. 
We have analyzed $\lambdas(m_T)$ data from $\sqrtsnn=200\ \GeV$ central Au+Au reactions measured at the BNL Relativistic Heavy Ion Collider (RHIC), using extensive Monte Carlo simulations based on six popular models for hadronic multiplicities.
Based on the combined STAR and PHENIX data set, and on various systematic investigations of resonance 
multiplicities and model parameters, we conclude that in $\sqrtsnn=200\ \GeV$ central Au+Au reactions 
the mass of the $\etap$ meson is reduced by $\Delta\meps>200\ \MeV$, at the 99.9\% confidence level in the 
considered model class. 
Such a significant \etap mass modification may indicate the restoration of the \UA symmetry in a hot and dense hadronic matter and the return of the ninth ``prodigal'' Goldstone boson.
A similar analysis of NA44 S+Pb data at top CERN Super Proton Synchrotron (SPS) energies showed no 
significant in-medium $\etap$ mass modification.
\end{abstract}

	

\maketitle

\section{\label{sec:intro}Introduction}

In terms of the quark model, one can observe a spontaneous symmetry breaking of the approximate $SU(3)_L \times SU(3)_R$ symmetry, resulting in nine pseudo-Goldstone bosons, that are usually associated with the light mesons formed as $u, d, s$ quark-antiquark bound states. This  na\"ive picture is, however, complicated by the fact that the $\etap$ meson has a large mass of  the order of 1 \GeV.
As early as 1970, Kobayashi and Maskawa concluded that the large mass of the \etap meson (formerly known as $X$) is a serious problem that is difficult to understand in a chiral $SU(3)_L \times SU(3)_R$ model with an explicit symmetry breaking term between singlet and octet states~\cite{Kobayashi:1970ji}. They found that the existence of an effective six-quark determinantial vertex is necessary. As shown by 't~Hooft in 1976, this vertex is contained in instanton-induced quark interactions~\cite{'tHooft:1976fv}. An interesting aspect of this Kobayashi-Maskawa-'t Hooft or KMT term~\cite{Kunihiro:2009ds} is that it can give rise to a flavor mixing in the scalar as well as in the pseudoscalar channels. 
The coupling between the pseudoscalar singlet and octet states $\eta_0$ and $\eta_8$ arises both from the $SU(3)_V$ breaking and the anomaly terms, assuming isospin symmetry. 
The physical $\eta$ and \etap mesons are given by the mixing of the $\eta_8$ and $\eta_0$ modes, and the mass of the $\eta_0$ singlet state turns out to be sensitive to the strength of the KMT vertex.
An explicit calculation for the general case gives the mixing angle $\theta_\eta(m_\eta^2)=-20.9^\circ$~\cite{Kunihiro:2009ds}. 

The $\eta$ and $\etap$ mesons change their masses as a function of the temperature $T$, due to both the $T$ dependence of the quark condensate, and the possible decrease of the KMT coupling constant with increasing $T$. The mixing angle $\theta_\eta$ also becomes $T$ dependent: as the temperature increases, mixing between $\eta$ and $\etap$ approximates the ideal one, and the $\eta_0$ component in the physical $\etap$ decreases.
On the other hand, with increasing $T$, the $\eta_0$ tends to play the role of the ninth Nambu-Goldstone boson of the $SU(3)_L \times SU(3)_R \times U_A(1)$ symmetry, and loses its mass rapidly.
At low temperatures, the $U_A(1)$ part of the symmetry is broken by instantons, invoking distinct vacuum states. Tunneling between these vacuum states is only possible ``at a cost'', giving extra mass to the \etap meson. However, as the transition amplitude is dependent on the strong coupling constant $\alpha_s$, it follows that the effect of instantons rapidly decreases with increasing energy density. 
This is an effective restoration of the $U_A(1)$ symmetry at finite $T$, first suggested in Ref.~\cite{Pisarski:1983ms}. Thus in high energy heavy ion collisions, where a hot and dense medium is created, the $U_A(1)$ symmetry may temporarily be restored~\cite{kunihiro,kapusta,huang}.

In Refs.~\cite{Vertesi:2009ca,prl} we reported the first observation of a significant reduction of the \etap mass, based on an analysis of PHENIX and STAR data~\cite{phnxpub,starpub} from \sqrtsnn=200 GeV central Au+Au collisions at the BNL Relativistic Heavy Ion Collider (RHIC). The subject of the present manuscript is to detail and systematically explore these experimental signatures of a partial $U_A(1)$ symmetry restoration at RHIC.

In thermal models, the production cross sections of the light mesons are exponentially suppressed by the mass. Since the \etap mesons are heavy, by default one expects the number of \etap mesons to be about two orders of magnitude less than the number of pions. However, as a consequence of the mass reduction, this suppression would be moderated, and the $\etap$ mesons would show up in an enhanced number. Once produced, the \etap is expected to be decoupled from other hadronic matter, since its annihilation cross section is very small. At the same time, the low-\PT \etap mesons are trapped in the medium due to energy conservation reasons, forming a condensate. Due to expansion, the medium cools down and freezes out, emitting asymptotic particles that propagate in $T=0$ vacuum. By this time, the \etap mesons regain their original mass, hence the enhancement will mostly appear at low $p_T$~\cite{kunihiro,kapusta,huang}.
It is to be noted that the \etap lifetime is much longer than the lifetime of the hot and dense medium, therefore a direct observation of the mass shift seems to be extremely difficult.

A promising channel of observation is the dileptonic decay $\etap\rightarrow\ell^+\ell^-\gamma$, because a low-\PT \etap enhancement would give extra lepton pairs to the low invariant mass region. The paper of Kapusta, Kharzeev and McLerran on the return of the prodigal Goldstone boson~\cite{kapusta} was in fact motivated by the dilepton enhancement seen in CERN Super Proton Synchrotron (SPS) $E_{Lab}=200A~\GeV/c$ laboratory energies in S+Pb reactions.
Recent interpretations of CERES~\cite{ceresll} and NA60 data~\cite{na60ll} indicate that the approach
to a chiral symmetry restored state could proceed through resonance broadening and eventually 
subsequent melting, rather than by dropping masses or mass dependency or mass degeneracy between 
chiral partners~\cite{Tserruya:2006ht}.
Recent PHENIX findings also show a definite excess in the $m_{\mathrm{e}^+\mathrm{e}^-}\lesssim 1\ \GeV$
dielectron invariant mass region in $\sqrtsnn=200\ \GeV$ Au+Au collisions~\cite{Afanasiev:2007xw}. 
Unlike at lower beam energies, in this case the contribution from a hot hadronic phase without mass 
shifts seems to be insufficient to account for the enhancement seen in the data~\cite{Drees:2009xy}.

In the present work we search for the \etap mass modification and the related restoration of the 
chiral symmetry in a different channel, using already published like-sign Bose--Einstein correlation (BEC) measurements as proposed by Ref.~\cite{vance}. We report on a detailed systematic study of the presently available published data sets. Future high precision data points in the low-\PT region will help to reduce the uncertainties of the measurements, and hopefully will provide a more precise estimation of the \etap mass reduction, compared to our current analysis.

\section{\label{sec:corr}Bose--Einstein Correlations}

Correlations between pions carry important information about the space-time structure of the medium 
created in heavy ion collisions. The widths of $\eta$ and \etap are  
$\Gamma_\eta=1.30\pm0.07\ \mathrm{keV}$ and $\Gamma_\etap=204\pm15\ \mathrm{keV}$, corresponding to large decay times: they produce pions at $c\tau_\etap\cong967\ \mathrm{fm}$ and $c\tau_\eta\cong152000\ \mathrm{fm}$, 
which are huge compared to the characteristic HBT radii of 4--6 fm.
Among the decay channels of the \etap, the $\etap\rightarrow\eta\pip\pim$ channel has the largest branching
ratio of about 45\%. Furthermore, the $\eta$ mesons decay into charged pions: The $\eta\rightarrow\pip\pim\pi^0$ and $\eta\rightarrow\pip\pim\gamma$ processes together have a branching ratio of approximately 27\%~\cite{pdg}.

The Bose--Einstein correlation function of pion-pion pairs can be expressed in terms of the relative and the mean 
four-momenta, $\Delta k \equiv \Delta k^{\mu} = p_1^{\mu}-p_2^{\mu}$ and $K \equiv K^{\mu} =(p^{\mu}_1+p^{\mu}_2)/2$, respectively:
\begin{equation}\label{eq:corr}
C(\Delta k,K)=\frac{N_2(\p{1},\p{2})}{N_1(\p{1})N_1(\p{2})}\ ,
\end{equation}
where \p{1,2} are the three-momenta of particles 1 and 2, $N_1$ and $N_2$ are the one- and two-particle invariant momentum distribution functions.\footnote{The usual $m^2 = p^2 \equiv  p^\mu p_\mu = E^2-\p{}^2$
notation is used here, where $p^\mu=(E,\p{})$ and \p{} is the three-momentum.}

It has been shown that the source can be handled in the core--halo picture, where the ``core'' consists of primordially created pions and those ones coming from fast-decaying resonances,
while the other pions, coming from more slowly decaying resonances, make the ``halo'' \cite{Csorgo:1994in,Bolz:1992ye}. The core region is resolvable by BEC, while the halo is not.\footnote{This has been tested numerically in~\cite{Nickerson:1997js}.} 
In fact, the core--halo separation always depends on the experimental two-track resolution. For example in PHENIX and STAR, two tracks are separable with a momentum difference of $\delta Q$ larger than 4 to 5 MeV, corresponding to a spatial separability $\delta x$ that dies off at about 40--50 fm~\cite{Adler:2006as,Sumbera:2007ir}. 
Long tails extending to this region were recently observed by the PHENIX and NA49 collaborations at RHIC and CERN SPS energies~\cite{Adler:2006as,Alt:2008fq}. Note that these long tails are also seen in kaon imaging, while the bulk production is well described with characteristic scales of 4--6 fm~\cite{Afanasiev:2009ii}.

In the core--halo picture~\cite{Csorgo:1994in,Bolz:1992ye} the correlation function can be measured with the so called extrapolated intercept parameter \lambdas as
\begin{eqnarray}\label{eq:clams}
C(\Delta k,K)=1+\lambdas R_c \ ,
\end{eqnarray}
where $R_c$ is defined by the Fourier transform of the one-pion emission function $S_c(x,K)$ of the core as
\begin{eqnarray}\label{eq:rc}
R_c &=& \frac{|\tilde{S_c}(\Delta k,K)|^2}{|\tilde{S_c}(0,K)|^2} \ , \nonumber \\
\tilde{S_c}(\Delta k, K) &=& \int d^4 x {S_c}(x,K)e^{i \Delta k x} \ ,
\end{eqnarray}
and the intercept parameter \lambdas{} is derived from the \textit{extrapolation} of the correlation value to $\Delta k=0$, as $\lambdas=C(0,K)-1$. However, this extrapolation does not include the correlation between halo--halo and core--halo particle pairs, supposed to be unresolvable by our detectors.\footnote{For fully thermal particle emitting sources, the exact value of intercept is $\lambda_{xct}=1$. Generally, $\lambda_{xct}>\lambdas$.}

The intercept parameter can also be expressed directly with the fraction of the core pions to the total number of pions:
\begin{equation}\label{eq:lams}
\lambdas(\mT)=\left(\frac{\NmT{\pip}{core}}{\NmT{\pip}{core}+\NmT{\pip}{halo}} \right)^2 \ ,
\end{equation}
where 
\begin{equation}\label{eq:mtcmp}
\NmT{\pip}{halo}=\NmT{\pip}{\omepip}+\NmT{\pip}{\etappip}+\NmT{\pip}{\etapip}+\NmT{\pip}{\kspip}\ ,
\end{equation}
%
and it is per se sensitive to the core vs.\ halo ratio.
(In the above formulae $\mT=\sqrt{m^2+\PT^2}$ denotes the transverse mass of the \pip{}s, and the $\NmT{\pip}{}$'s are the \mT{} distributions of the corresponding decays.)
If one does a one-dimensional investigation characterized by the invariant momentum difference $q=\sqrt{(k_1-k_2)^2}$, and assuming a Gaussian approximation of the source, one will have Eq.~(\ref{eq:clams}) in the form
\begin{eqnarray}\label{eq:cgaus}
C(q)=1+\lambda e^{-|qR|^2} & .
\end{eqnarray}
In this case $\lambda\equiv\lambdas$.

As \lambdas is defined in Eq.~(\ref{eq:clams}) as an {\it extrapolation} of the experimental data to $\Delta k^\mu\rightarrow 0$, it is essential that it does not depend strongly on the method of the extrapolation. The Gaussian 
assumption for the shape of the source is a model hypothesis that has been shown not to be 
precise~\cite{Adler:2006as}. Instead, one can choose an experimental procedure independent of 
theoretical assumptions, and get a description of arbitrary precision with the Edgeworth expansion
by completing the Gauss shape with an infinite series in the space of Hermite polynomials \cite{csh}. In this case,
\begin{eqnarray}\label{eq:cedge}
C(q)=\mathcal{N}\left[ 1+\lambda_E e^{-q^2 R_E^2} \left( 1+\sum_{n=3}^\infty \frac{\kappa_n}{n!} H_n(\sqrt{2}qR_E) \right) \right],
\end{eqnarray}
where $\mathcal{N}$ is an overall normalization parameter, the $\kappa_n$ is a weight factor to the $n$th Hermite polynomial
\begin{eqnarray}\label{eq:herm}
H_n(t)=e^{t^2/2}\left( -\frac{d}{dt} \right)^n e^{-t^2/2}& ,
\end{eqnarray}
and $R_E$ and $\lambda_E$ are the Edgeworth radius and intercept parameters. In practice, an expansion up to the sixth order is sufficient~\cite{starpub}. Note that, as the even order Hermite polynomials have non-vanishing values at $t=0$, $\lambda_E$ is different from $\lambdas$ here:
\begin{eqnarray}\label{eq:lame}
\lambdas=\lambda_E \left[1+\frac{\kappa_4}{8}-\frac{\kappa_6}{48}+... \right] & .
\end{eqnarray}

A straightforward generalization of these approximations to the three (out, side, long) dimension case is given in \cite{Eggers:2006uq}.

In the following we will show that the phenomenon of a partial \UA{} restoration and a connected mass reduction of the $\etap$ meson is able to explain the behavior of $\lambdas$ measured in RHIC experiments.
A previous analysis, based on the same idea, was carried out on NA44 data, and showed that a reduction of the $\metap$ mass should result in a dip of the $\lambdas(\mT)$ at low-\mT{} values. However, no significant signal of such a mass modification was seen in $E_{lab}=200A\ \mathrm{GeV}$ S+Pb collisions at CERN SPS (Fig.~1 of \cite{vance}).
 
\section{\label{sec:datasets}Data sets}

\noindent
In contrast to the SPS data, a low-\mT dip of the $\lambdas(\mT)$ was measured in $\sqrt{s_{NN}}=200\ \GeV$ Au+Au collisions at RHIC by both the STAR and the PHENIX experiments~\cite{phnxpub,starpub}. The $\lambdas(\mT)$ points were extracted from the measured correlation function of like-sign pions, using different methods. 
A comparison of the measurements to the FRITIOF~\cite{frirqmd} calculation including variation of the \etap mass was presented in Fig.~4. of Ref.~\cite{Csanad:2005nr}.
It is important to note that, although the Gaussian fit typically has 1--2\% error on the $\lambda$ value, this does not reflect the real error on the intercept caused by the extrapolation. For example, exponential fits yield larger values for $\lambda$ than the Gaussian fits do, and the difference between $\lambda_{Gauss}$ and $\lambda_{exponential}$ is larger than several (sometimes more than 5) standard deviations~\cite{Hegyi:1992vc}. A reasonable range of errors can be estimated with the help of an Edgeworth fit~\cite{Csanad:2005nr}.
In the detailed presentation of Ref.~\cite{Csanad:2005nr} it was suggested to utilize the normalized \lamfrac quantity in order to remove sensitivity to the extrapolation technique, and to reduce other systematic errors. Note that $\lambda_*^{max}$ is defined as
the value of the extrapolated intercept parameter in an $\mT$ region where its value is saturated; according to simulations as well as 
the presently used PHENIX and STAR measurements, this corresponds to 0.5 GeV $\le \mT \le $ 0.7 GeV region.%
\footnote{%
  Here $\lambdas^\mathrm{max}$ is the $\lambdas(\mT)$ value taken at $\mT=0.7\ \GeV$, with the exception of the STAR data, where the data point at the highest $\mT=0.55\ \GeV$ is considered. Note that the \mT dependency of the  $\lambdas(\mT)$ measurements in the 0.5--0.7 GeV region is very weak.
}
One should also note that the $\lambdas(\mT)$ data may depend on the goodness of the particle identification too:
other particles misidentified as pions will reduce correlation and will push the measured $\lambdas(\mT)$ data down.

The PHENIX $\lambdas(\mT)$ data set was derived using the Bowler-Sinyukov method \cite{sinyukov}, while the STAR data set was obtained from a sixth order Edgeworth fit. Here we apply the method of Ref.~\cite{csh} using the values and errors of the Gaussian $\lambda$ and the $\kappa_{i,n}$ Edgeworth fit parameters taken from \cite{starpub} in order to compute the sixth order $\lambdas$ values using Eq.~(\ref{eq:lame}). We use complete error propagation in order to determine the corresponding uncertanities of $\lambdas$.
Although the Gaussian $\lambda$ and the Edgeworth $\lambda_E$ parameters are significantly different, the extrapolated intercept parameter \lambdas is similar to $\lambda$ within errors (see Table~\ref{tab:starlam} and Fig.~\ref{fig:starlam} in Appendix \ref{sec:starlam}). 
However, the error bars on $\lambdas$ are significantly larger than the na\"ive errors on $\lambda$. 
As emphasized before, the Edgeworth $\lambdas$ gives a more realistic estimation of both the value and the error of the intercept parameter, hence of the core-halo ratio.

\begin{figure}[h!tbp]
\begin{center}
\includegraphics[width=.50\linewidth]{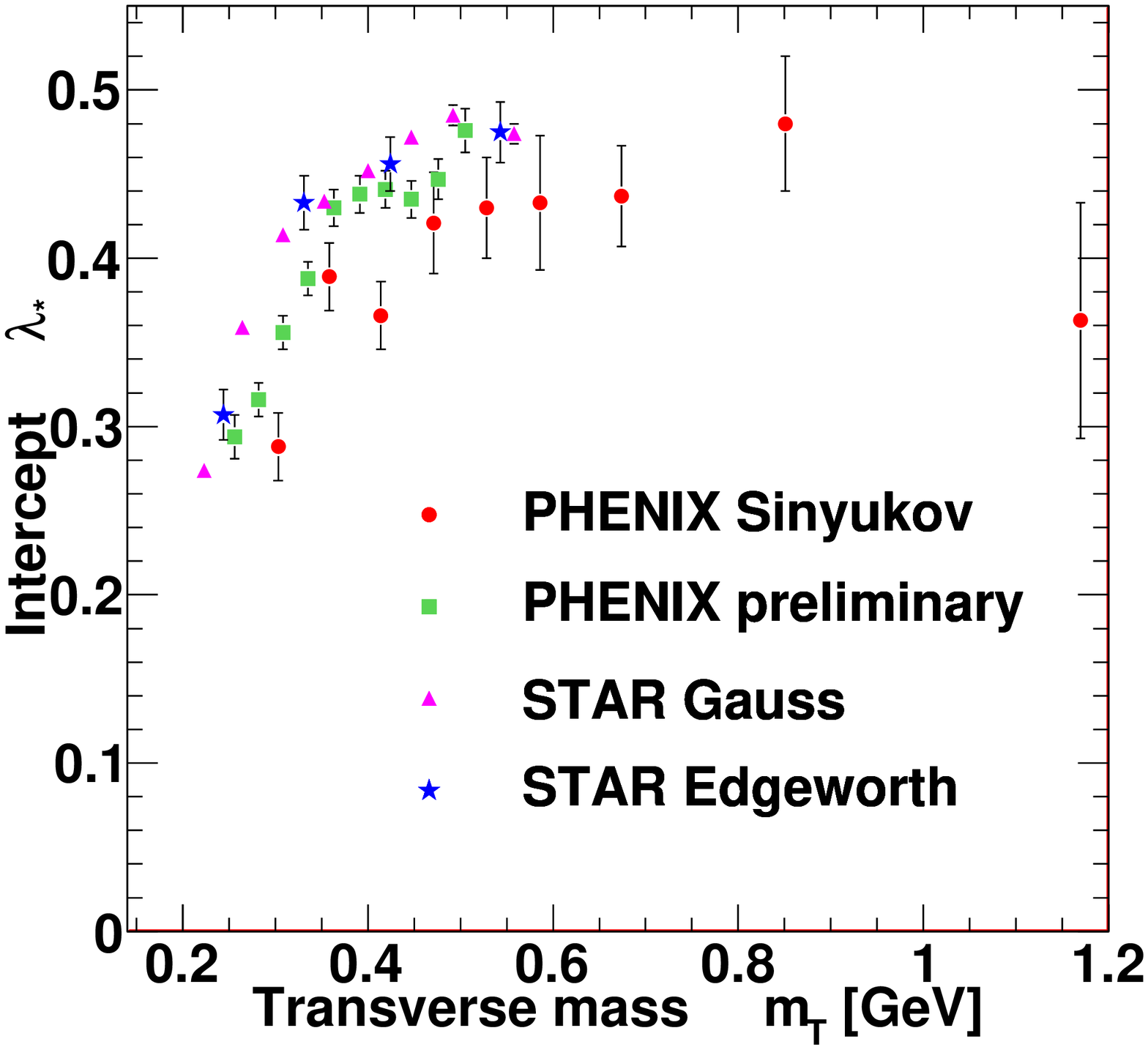}%
\includegraphics[width=.50\linewidth]{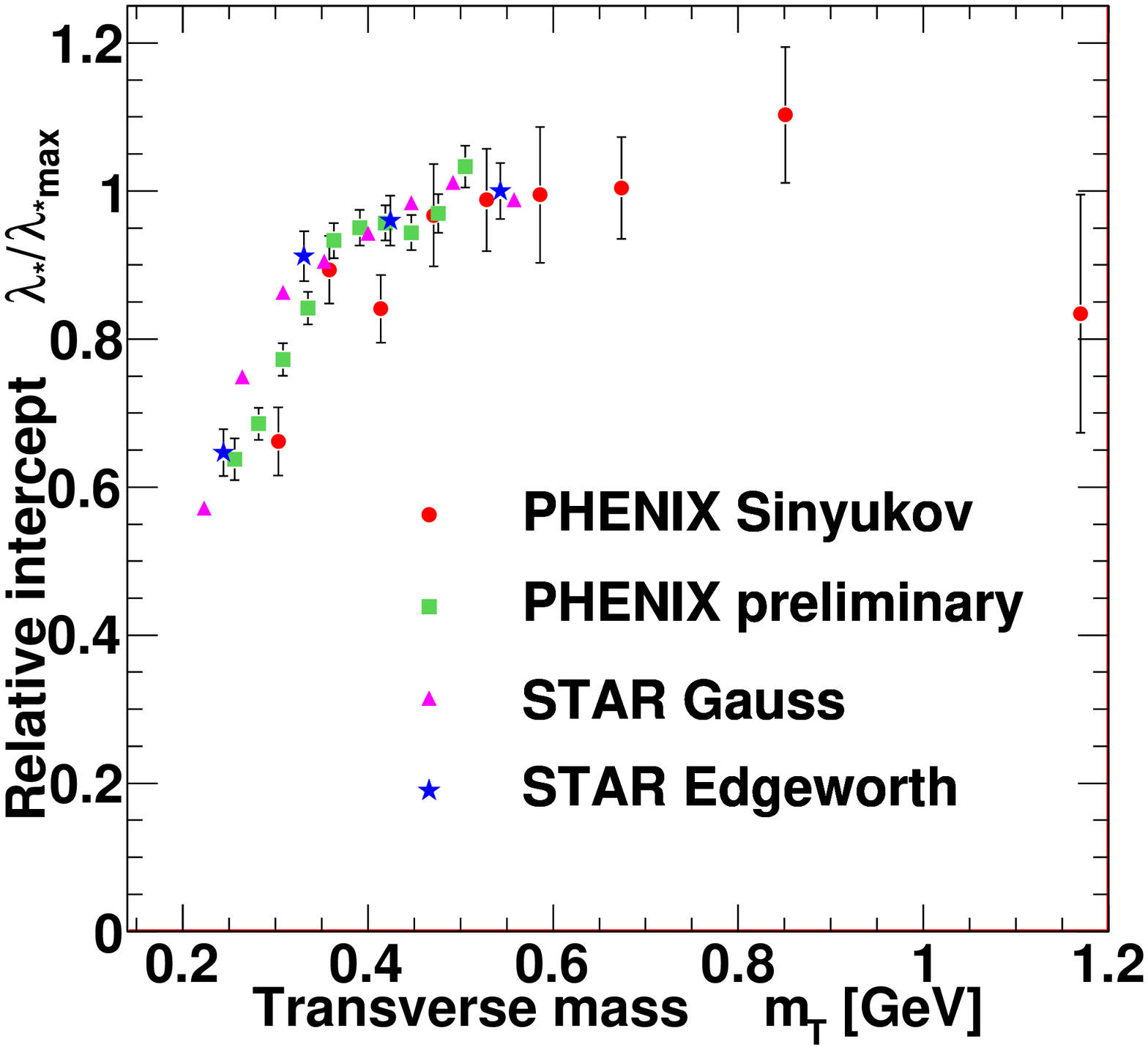}
\end{center}
\caption{{\it (Color online)} Data sets of $\lambdas(\mT)$ {\it(left)} and \lamfrac {\it(right)} from RHIC 
$\sqrt{s_{NN}}=200\ \GeV$ like-sign pion correlation measurements.
Note that the errors on the STAR Gaussian data set (Fig.~14 of Ref.~\cite{starpub}) are in the order of 0.001~MeV (not all systematics included), and one of the PHENIX data sets has only preliminary errors~\cite{Csanad:2005nr}. 
Hence we are left with the remaining two data sets when evaluating the systematic errors on \meps: The published PHENIX data set of Ref.~\cite{phnxpub}, and the STAR data set calculated from the data shown in Fig.~13 of Ref.~\cite{starpub} using the method of Appendix~\ref{sec:starlam}.}
\label{fig:datasets}
\end{figure}

Another data set on \lambdas from STAR was obtained with a Gaussian fit using the Bowler-Sinyukov method, and it shows a good agreement with the STAR Edgeworth data (Fig.~14 of \cite{starpub}). We do not use this data set in our analysis, since the error bars do not include all the relevant systematic effects, being in the order of $\delta\lambda\sim 0.001$, which is way smaller than the systematic error coming from the choice of the extrapolation. PHENIX preliminary data \cite{Csanad:2005nr} are also shown, for comparison purposes only, as their systematic errors are not yet finalized. The data sets for \lambdas are detailed in Fig.~\ref{fig:datasets} before and after normalizing with \lambdasmax. Each of these data sets indicate the dip of \lambdas in the low-\PT region.

Given that we are interested in obtaining final errors that include all relevant systematic effects, we decided to 
analyze simultaneously the PHENIX final Sinyukov-corrected \lambdas data set together 
with the STAR Edgeworth \lambdas data set, both normalized to their \lambdasmax values. One can 
note that these two data sets are not in perfect overlap with each other, although they are consistent 
within their errors, and it is possible to fit both data sets simultaneously with good confidence levels. 
The Gaussian $\lambda(\mT)$ data set of STAR has errors that apparently do not include systematics from the Gaussian ansatz, 
thus a quantitative comparison to our model was not reasonable. 
However, the best fits to the other data sets qualitatively agree with the Gaussian $\lambda(\mT)$ points.
We have also checked that separate analysis of the STAR and PHENIX data sets yields results which are consistent
with the presented results, as part of the systematic studies. However, the combined PHENIX and STAR data set provided a more precise estimate for the allowed regions of the model parameters.

%
The difference between the selected PHENIX and STAR data sets possibly reflects 
the systematic error from different experimental conditions, for example, PHENIX data were measured in the 0-30\% centrality class,
while STAR Edgeworth results were published for the 0--5\% centrality selection, and the particle identification
in the two experiments is also different. These differences resulted in a systematic uncertainty of our analysis, too,
however, as we shall demonstrate at the end, these systematic errors are still of the order of the statistical uncertainties
and the dominant error in estimation of the in-medium modified mass of the $\etap$ mesons comes from the choice of the
resonance model and its parameters. The relative systematic error from the difference in centrality selection is estimated in
Appendix~\ref{sec:centrality} to be not larger than 9.8\%. 

\section{\label{sec:simulation}Modeling and simulation}

First simulations of the \UA restoration at SPS S+Pb collisions at $E_{lab}=200A\ \mathrm{GeV}$ bombarding energy had predicted a dip of $\lambdas(\mT)$ at \mT values below 0.25 \GeV{}~\cite{vance}. It was found that the depth of this dip was governed by the value of \meps as an input parameter for the simulations. In those simulations, the \etap mesons from the condensate had been assumed to have no transverse momenta at all, resulting in a very steep hole-like structure of the low-\mT part of the $\lambdas(\mT)$~\cite{vance}. Such an oversimplification is not adequate for the description of the dip in the RHIC data. To improve on it, we introduce here an effective thermal spectrum for the \etap mesons from the condensate, characterized by an inverse slope parameter \Binv. 
As we demonstrate below, this parameter \Binv controls the steepness of the dip of $\lambdas(m_T)$, 
while \meps controls its depth. Also, instead of relying on a given model of resonance production
(as on FRITIOF~\cite{frirqmd} in Ref.~\cite{vance}), here we 
utilize six different models, in order to estimate the systematic error related to the choice of
the theoretical model for resonance production. In particular, the six models we use here are ALCOR \cite{alcor}, FRITIOF \cite{frirqmd}, Kaneta {\it et al.}
\cite{kaneta}, Letessier {\it et al.} \cite{rafelski}, Stachel {\it et al.} \cite{stachel} and
UrQMD \cite{urqmd}. (See Appendix~\ref{sec:models} for more details on the individual models.)
We have also considered the AMPT 2.11 $\lambda(\mT)/\lambda^\mathrm{max}$ simulation with string melting~\cite{Lin:2004en} that report on a non-thermal scenario without an \etap mass modification. 
AMPT results show an interesting, although genuinely different dropping structure of $\lambda(\mT)$ at low \mT values. Comparison of this model to the data, as tersely overviewed in~\cite{prl}, 
indicate that AMPT cannot describe these data in a statistically acceptable manner, characterized by a $\chi^2/ndf=102/13$ corresponding to $\rm{CL}=6.8 \times 10^{-16}$. We attribute the behavior seen in AMPT to a lower effective $\langle u_T\rangle$ of the high mass halo resonances~\cite{prl,vance}.

In earlier simulations in~\cite{vance}, the resonance production was generated by an exponential spectrum 
$N(\mT)=A e^{-\mT/\Teff}$, the effective freeze-out temperature defined as $\Teff=\Tfo+m\uT^2$, with \Tfo and \uT being the freeze-out temperature and the average transverse flow, respectively.
This has also been generalized, and a polynomial prefactor had been introduced in order to achieve a more realistic description of the direct production of resonances. We have fixed \Tfo and \uT to
RHIC measurements~\cite{Adler:2003cb}. Thus the \mT distribution will follow the form of
\begin{equation}\label{eq:mtdist}
N(\mT)=C\mT^{\alpha}e^{-\mT/\Teff},
\end{equation}
where $C$ is a normalization constant, and $\alpha=1-d/2$, where $d$ is the number of spatial
dimensions of the expansion (hence $1\leq d \leq 3$ and $\alpha$ falls 
between $-1/2$ and $1/2$)~\cite{Csorgo:1995bi,Csorgo:1994in}.
The choice of $\alpha=0$ corresponds to the case of Ref.~\cite{vance}. 

In order to compare to RHIC data at mid-rapidity, we compute the effective intercept parameter
of the \pip{}--\pip{} correlations, $\lambdas(\mT)$, with the definition of Eq.~(\ref{eq:lams}). All the
contributions of Eq.~(\ref{eq:mtcmp}) to the $\mT$ distribution were simulated in the mid-rapidity region,
given by the pseudorapidity\footnote{In this work, pseudorapidity is denoted by $ y = 0.5 \ln[(|p| + p_z)/(|p|-p_z)]$.
In other manuscripts $y$ is usually denoted by the variable $\eta$ which in the current work is reserved for the 
pseudoscalar mesons  $\eta$ and $\etap$.} cut of $|y|<0.36$ of the PHENIX acceptance. In the systematic checks we took into account that
the STAR analysis~\cite{starpub} used a different pseudorapidity cut of $|y|<0.5$. 

For each model the fractions of the particles were computed by normalizing the multiplicities to the total density 
\begin{equation}\label{eq:rhotot}
\rho_{total}=\rho_{core}+\rho_{halo} 
, \quad
\rho_{halo}=\rho_\omega+\rho_\etap+\rho_\eta+\rho_{\ks}.
\end{equation}
The halo of pion production always had the same ingredients (from the decays of the $\omega$, $\etap$, $\eta$ and $\ks$)\footnote{Other long-lived resonances, such as the $\phi$ meson, are checked to give a negligible contribution to the \lamfrac ratio, which translates to less than 2\% uncertainty to the \meps.}, while the core was composed of all the other resonances that were available in each particular model.
The charged pion \mT spectra were obtained from a complete kinematic simulation of the decays above resonances using JETSET v7.4 \cite{Sjostrand:1995iq}. The estimated systematic error arising from assigning each $\omega$ decay product to the halo is given in Sec.~\ref{sec:results}.

The mechanism of the partial \UA{} restoration implies that the $\etap$ would have a decreased 
effective mass in the hot and dense medium \cite{vance}. The number of the created $\etap$ particles 
would then follow Eq.~(\ref{eq:mtdist}) with the modified mass and the freeze-out temperature respective 
to the $\etap$ mesons, and the fraction of $\etap$ mesons in the condensate is modeled with the
\begin{equation}\label{eq:prietamtdist}
f_\etap \equiv \frac{N_\etap^{*}}{N_\etap}=\left(\frac{\meps}{\metap}\right)^\alpha e^{\frac{\metap-\meps}{\Tcond}}
\end{equation}
formula, where the $T_{cond}$ is the temperature of the condensate. 

As the escaping $\etap$ bosons are regaining their mass, they must lose momentum in order to fulfill the principle 
of energy conservation, i.e., 
\begin{eqnarray}\label{eq:metap}
{m_{\etap}^*}^2+{p_{T,\etap}^{*}}^2={m_{\etap}}^2+{p_{T,\etap}}^2 & .
\end{eqnarray}
(In the above equation the quantities with an asterisk denote the properties of the in-medium \etap{}, while the 
ones without an asterisk refer to the free \etap. According to the kinematical setup of both PHENIX and STAR measurements~\cite{phnxpub,starpub}, the longitudinal component of the \etap momentum is considered to be negligible here.)
As a consequence, while \etap bosons with ${p_{T,\etap}^{*}} > \sqrt{ {m_{\etap}^*}^2-{m_{\etap}}^2 }$ 
will follow the above distribution (with the effective mass $m_{\etap}^*$ plugged in), the ones moving 
with a momentum less than this limit will be ``trapped'' in the medium until this medium is dissolved.
Afterwards, \etap{}-s from the condensate are given a random transverse momentum, following 
Maxwell-Boltzmann statistics with a characteristic temperature \Binv{}:
\begin{eqnarray}\label{eq:mb}
f(p_x,p_y)=\left( \frac{1}{2\pi\metap\Binv}\right)e^{-\dfrac{p_x^2+p_y^2}{2\metap{\Binv}}} \quad ,
\end{eqnarray}
with $p_x^2+p_y^2=p_{T,\etap}^2$. Note that $B$ is a systematically varied model parameter, and its best value is determined by the analysis of the $\lambdas(\mT)$ data.

\section{\label{sec:results}The \etap mass drop---results and systematics}

Out of the six parameters of the \etap spectrum and $\lambdas(\mT)$ simulation, the two most 
important ones are \Binv and \meps. They directly determine the shape of the observed dip 
(depth and width, respectively). These parameters were considered as ``fit variables,'' and a 
fine grid $\chi^2$ scan was carried out. The most probable values of the \meps\ and \Binv\ were 
looked for by the simulation of the $\lambdas(\mT)$ spectrum and then determining the $\chi^2$ 
from the fits to the data of the PHENIX and STAR measurements at each individual setup. 
The other four parameters were treated as ``constants,'' although their variation played a role 
when determining systematic errors.

As a default setup, the \pip freeze-out temperature was taken from PHENIX inverse slope parameter
fit\footnote{Note that the values for \Tfo and \uT 
only include the statistical errors. The systematics were taken into account in our systematic studies that involved a wide range of different predictions on \Tfo, $100\ \MeV\leq\Tfo\leq 177\ \MeV$.} to be $\Tfo=177.0\pm1.2$ MeV \cite{Adler:2003cb}, and the conservative assumption of 
$\Tcond=\Tfo$ was made.\footnote{The temperature of the condensate majorates the freeze-out temperature. As can be seen from our systematic checks, a freeze-out temperature that is lower than the condensate temperature will move \metap downward, i.e.,\ toward ``safety'' for our conclusion.}
The average transverse flow velocity was measured by PHENIX to be $\uT=0.48\pm 0.07$ (in relativistic units) 
\cite{Adler:2003cb}. This was cross-checked in our simulations with a $0.40\le\uT\le0.60$ scan with $\Delta\uT=0.01$ steps at the best value of \meps and \Binv, and it was found that the most probable values were around $\uT=0.50$, in agreement with the PHENIX measurement. The polynomial exponent $\alpha=0$ corresponds to a 2D expansion~\cite{Csorgo:1994in,Csorgo:1995bi}, that had been proved to give a good description of data, used by PHENIX when obtaining $\left\langle u_T \right\rangle$ and \Tfo. However the $\alpha=+1/2$ (1D) and $\alpha=-1/2$ (3D) cases were also examined as systematic checks. 

\begin{figure}[h!tbp]
\begin{center}
\includegraphics[width=.50\linewidth]{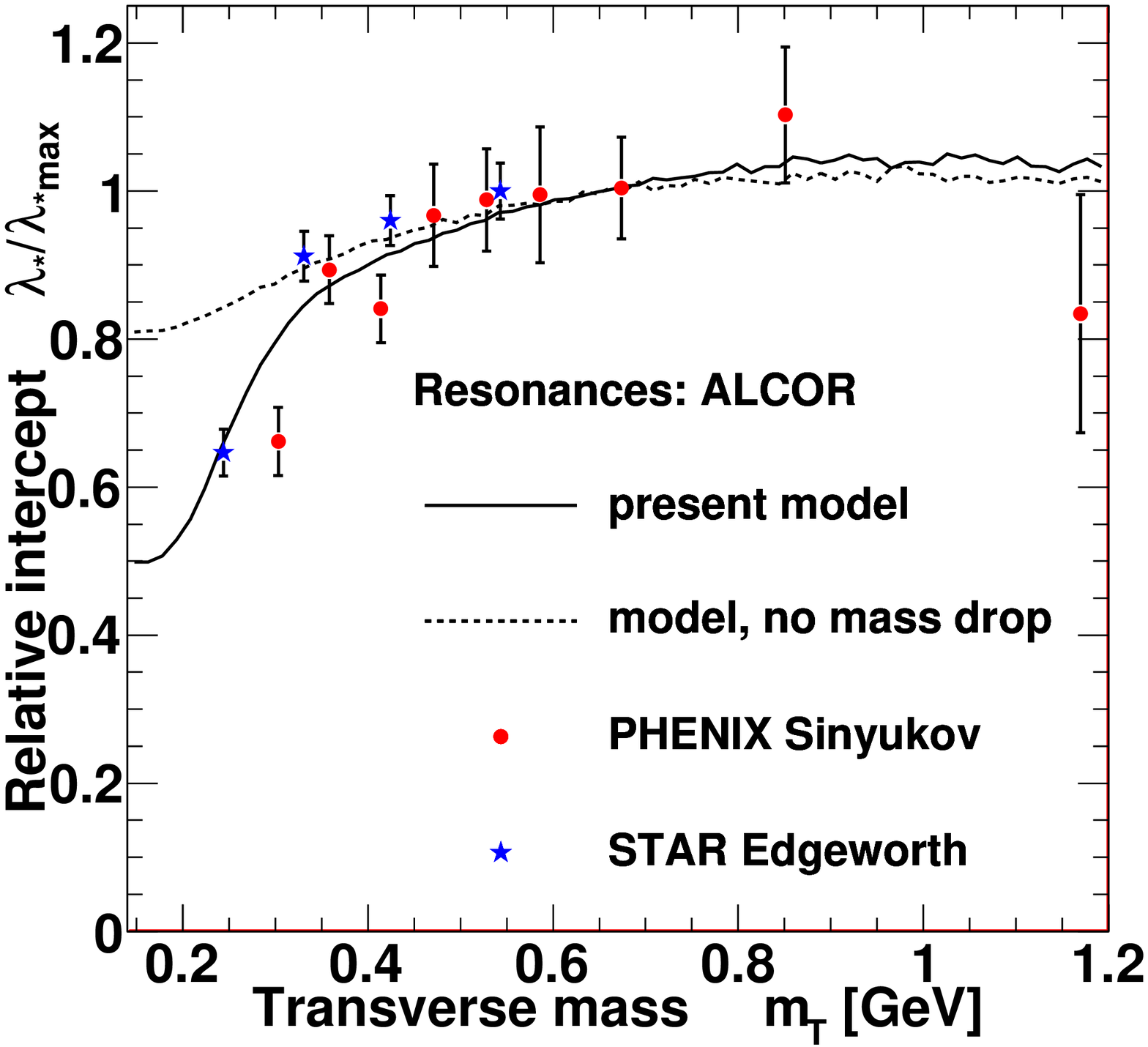}%
\includegraphics[width=.50\linewidth]{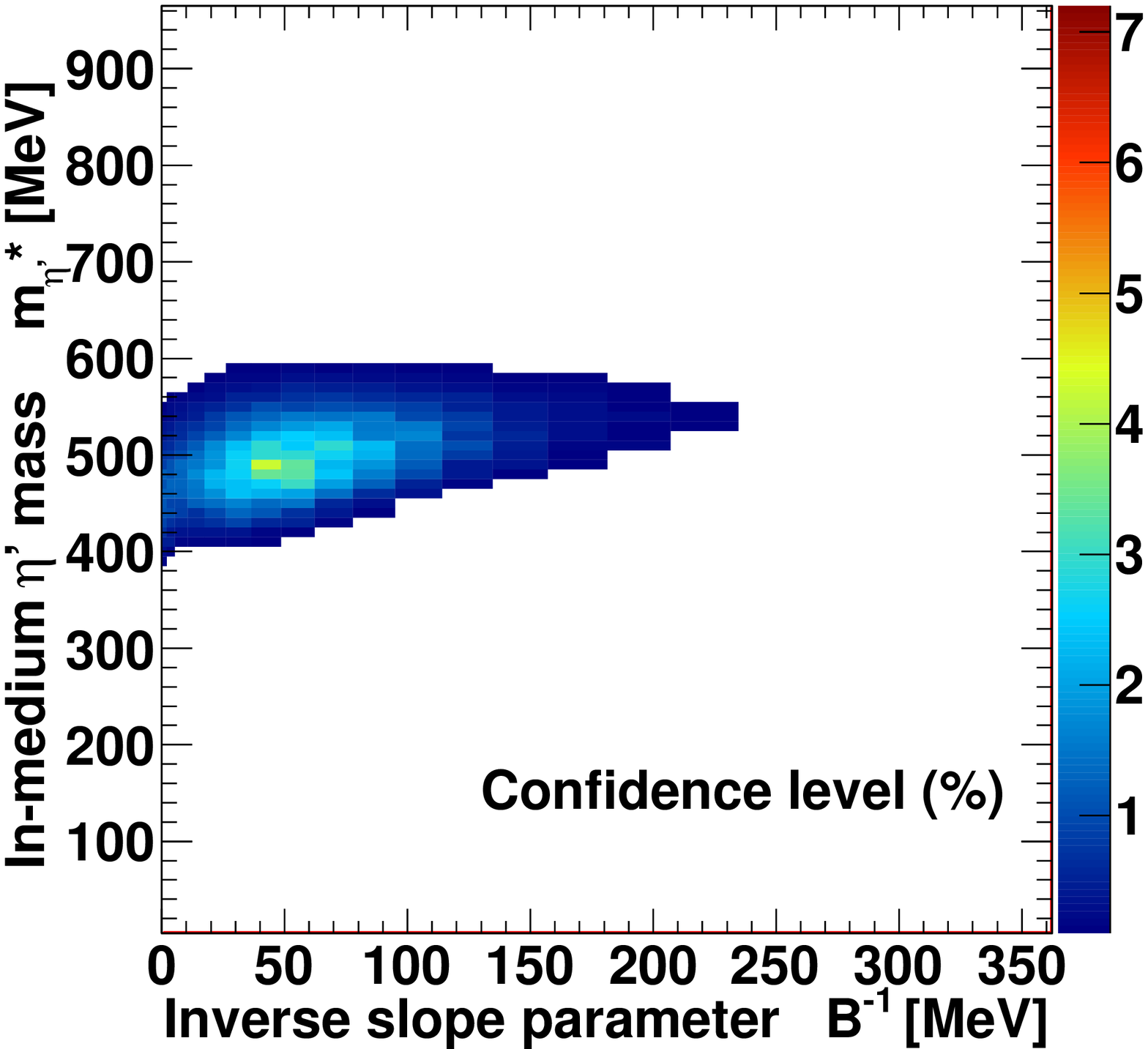}
\includegraphics[width=.50\linewidth]{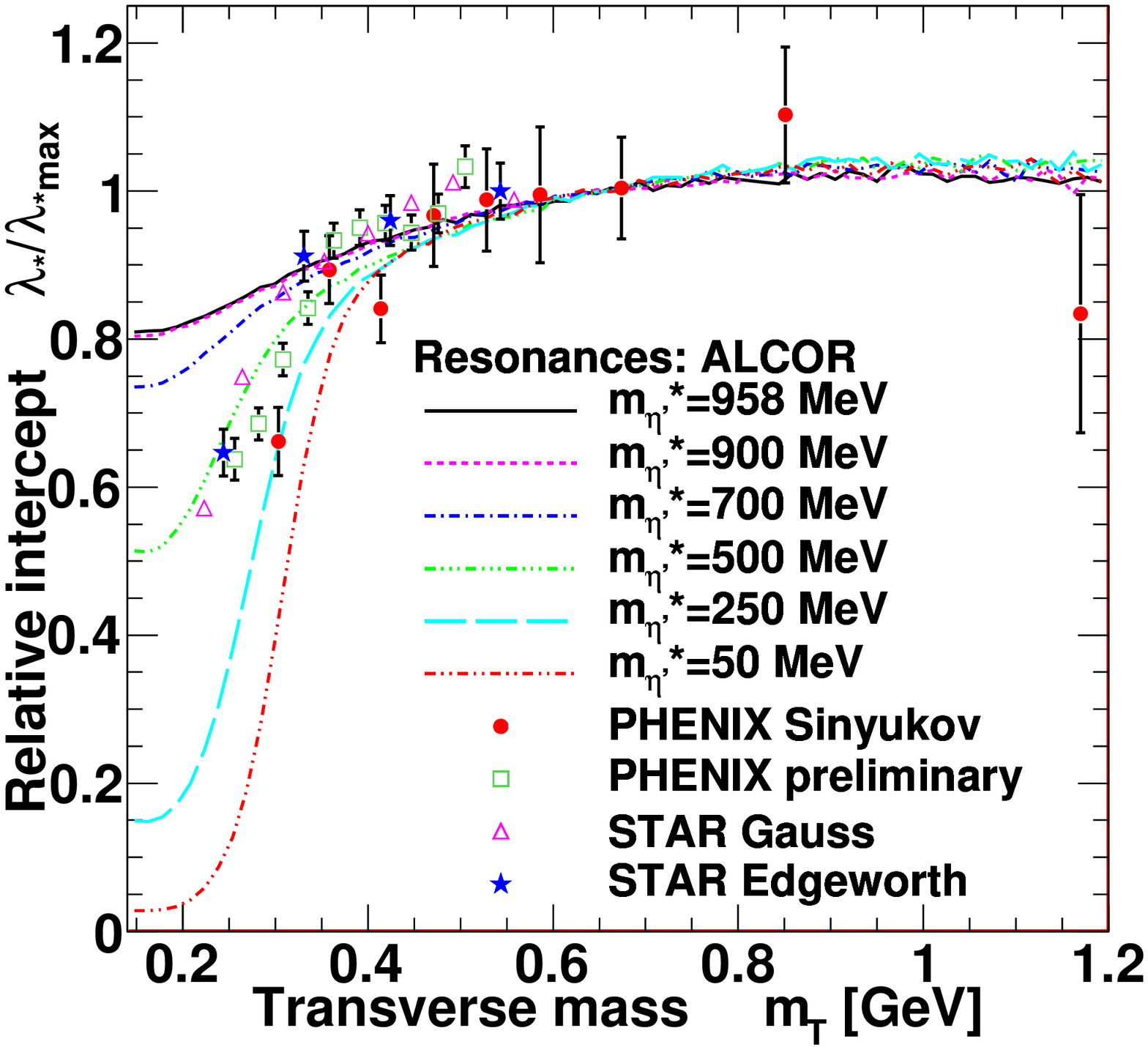}%
\includegraphics[width=.50\linewidth]{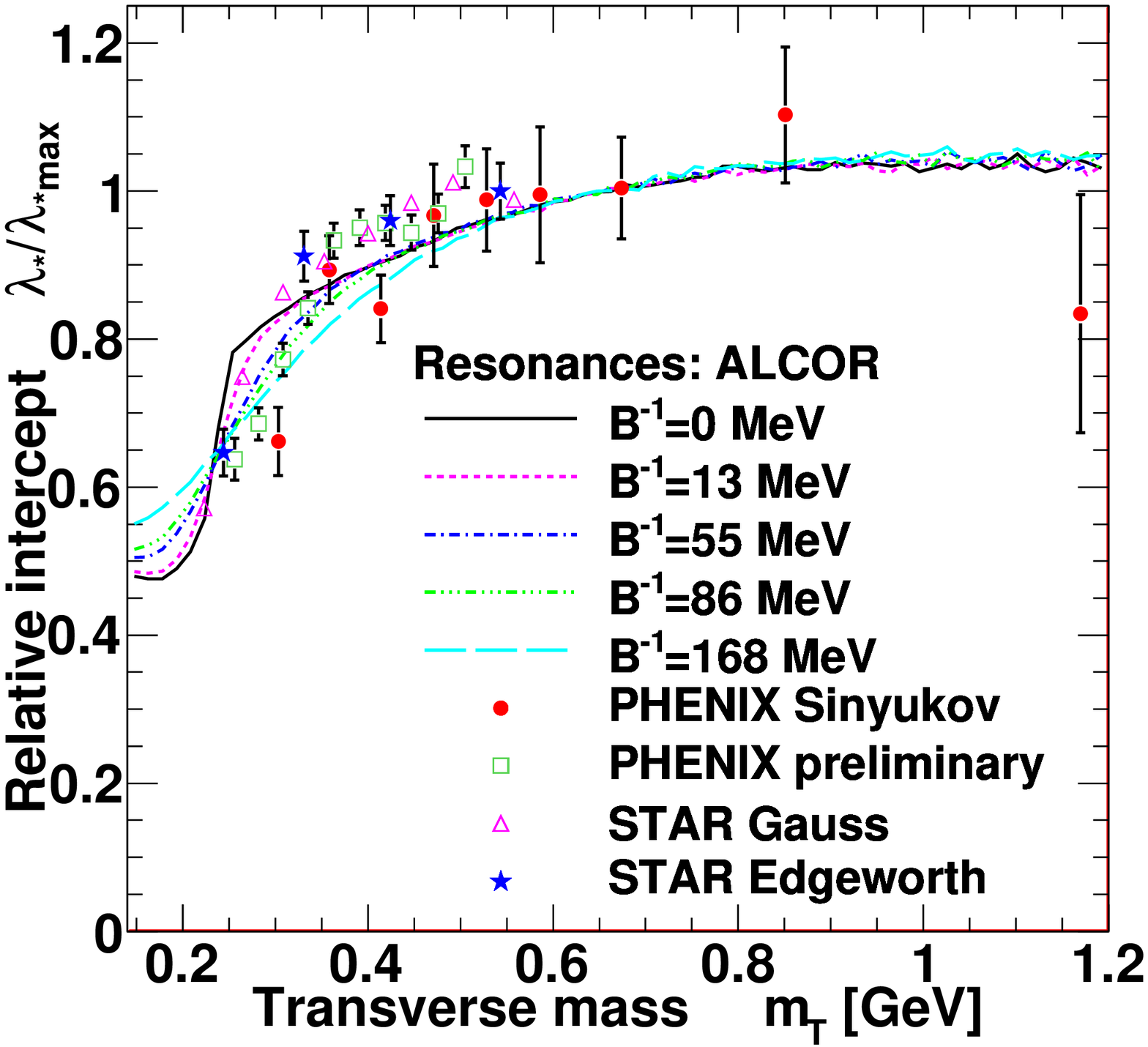}
\end{center}
\caption{{\it (Color online)}
{\it Top left:} \lamfrac\ spectrum of the PHENIX and STAR data points, in comparison with Monte Carlo simulations based on the ALCOR model \cite{alcor} at parameters from the best fit ($\Binv=42$ MeV, $\meps=490$ MeV), for $\alpha=0$, $\Tcond=177$ MeV, $\Tfo=177$ MeV and $\uT=0.48$.
{\it Top right:} Corresponding confidence level surface on the (\Binv{}, \metap) plain. Only the region with acceptable fits ($CL>0.1\%$) is shown.
{\it Bottom left:} \lamfrac\ spectra for different $\meps$ values, $\Binv$ fixed to its best fit value. 
{\it Bottom right:} \lamfrac\ spectra for different $\Binv$ values, $\meps$ fixed to its best fit value.
In the latter two plots the PHENIX preliminary and the STAR Sinyukov measurements are also shown for 
comparison (as explained in Sec.~\ref{sec:datasets}).
}
\label{fig:alcor}
\end{figure}

\begin{figure}[h!tbp]
\begin{center}
\includegraphics[width=.50\linewidth]{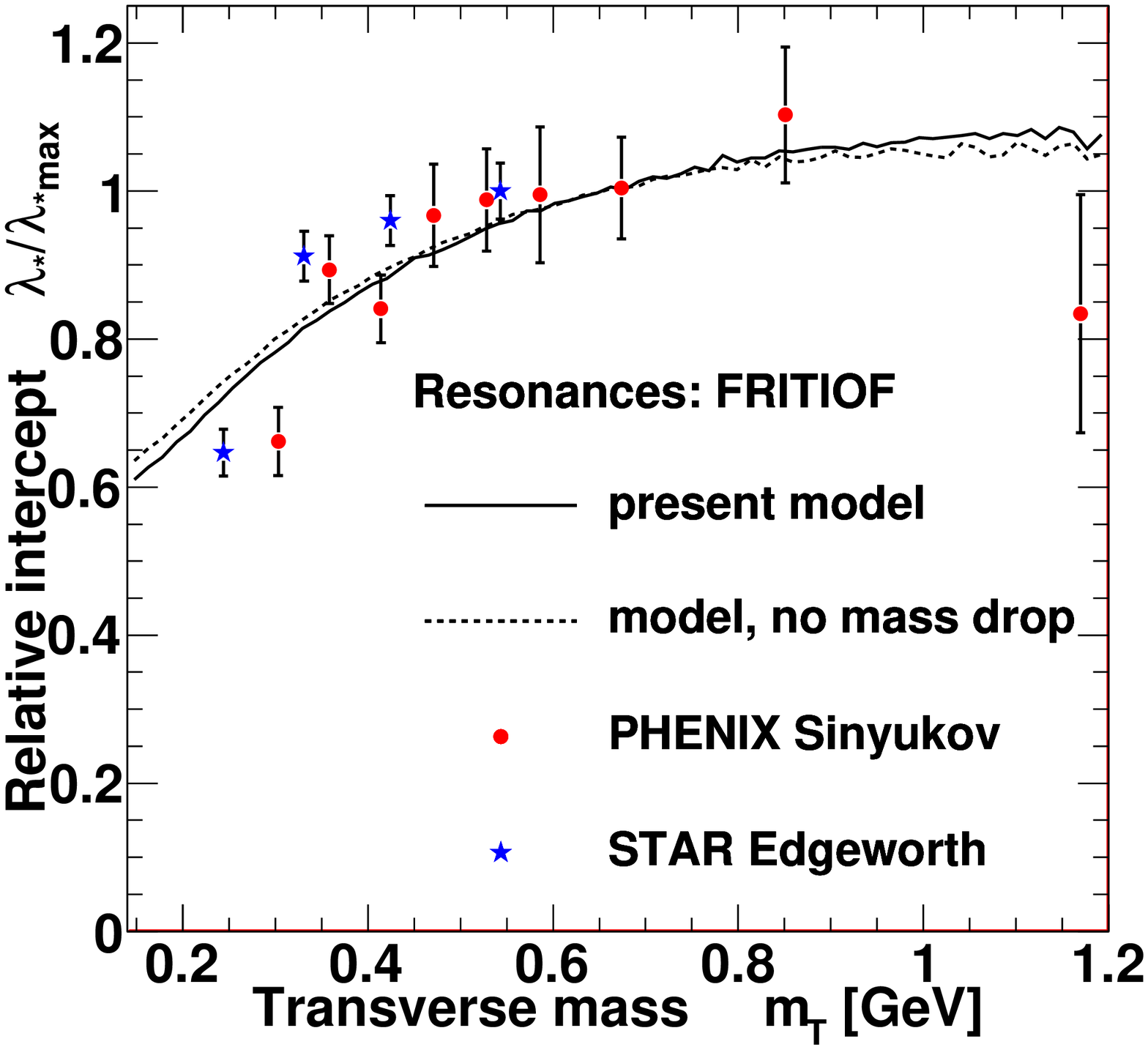}%
\includegraphics[width=.50\linewidth]{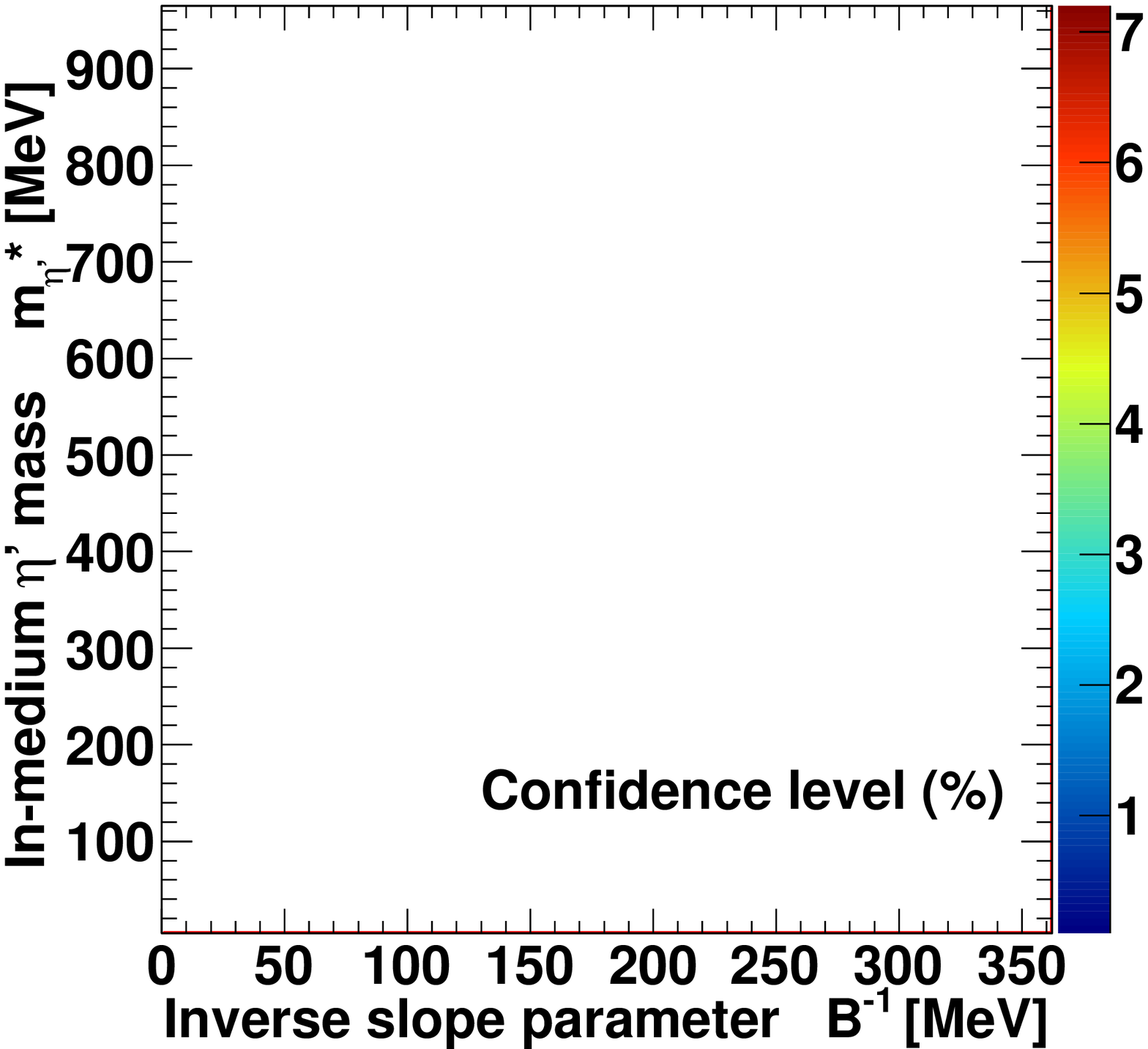}
\includegraphics[width=.50\linewidth]{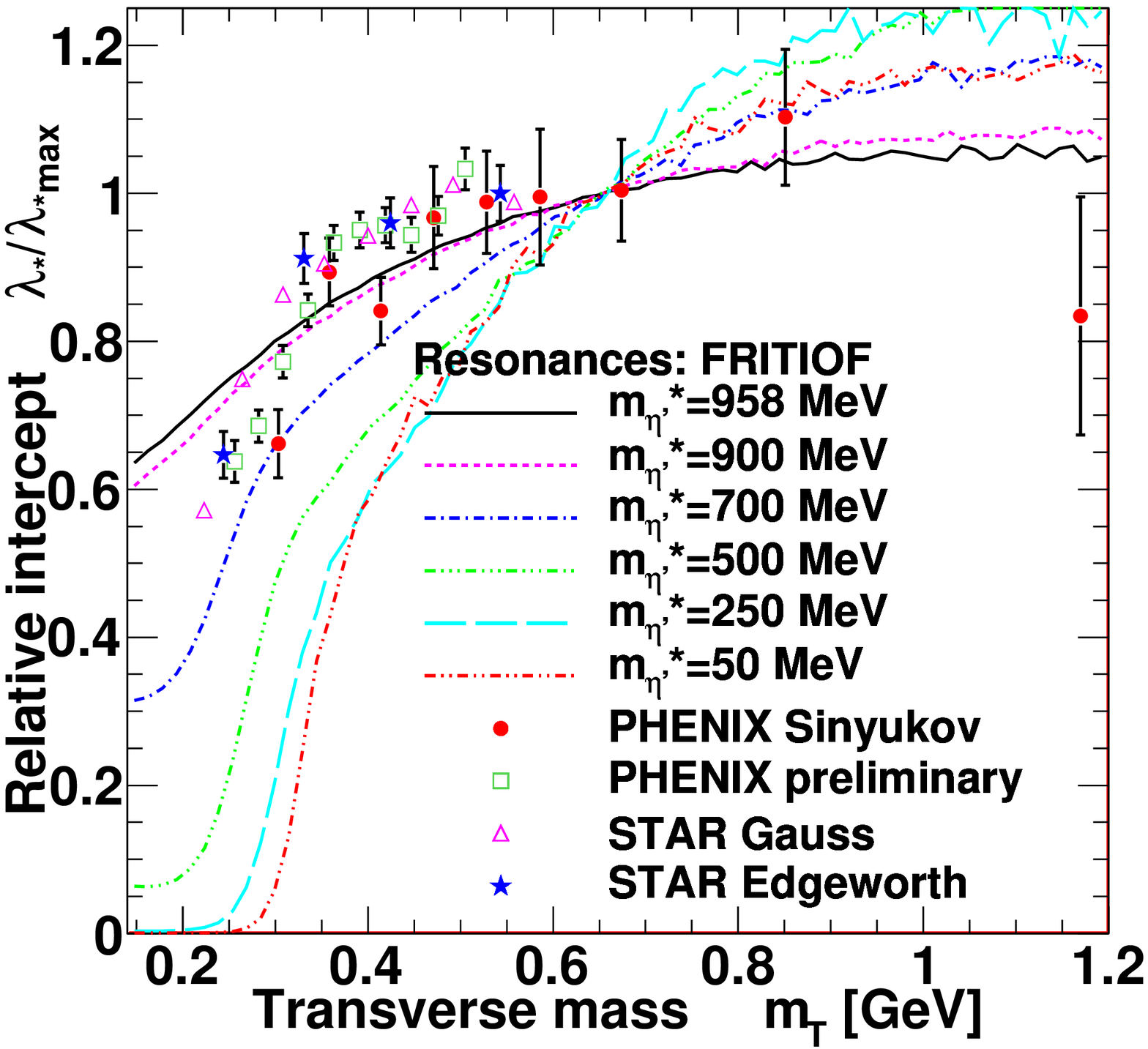}%
\includegraphics[width=.50\linewidth]{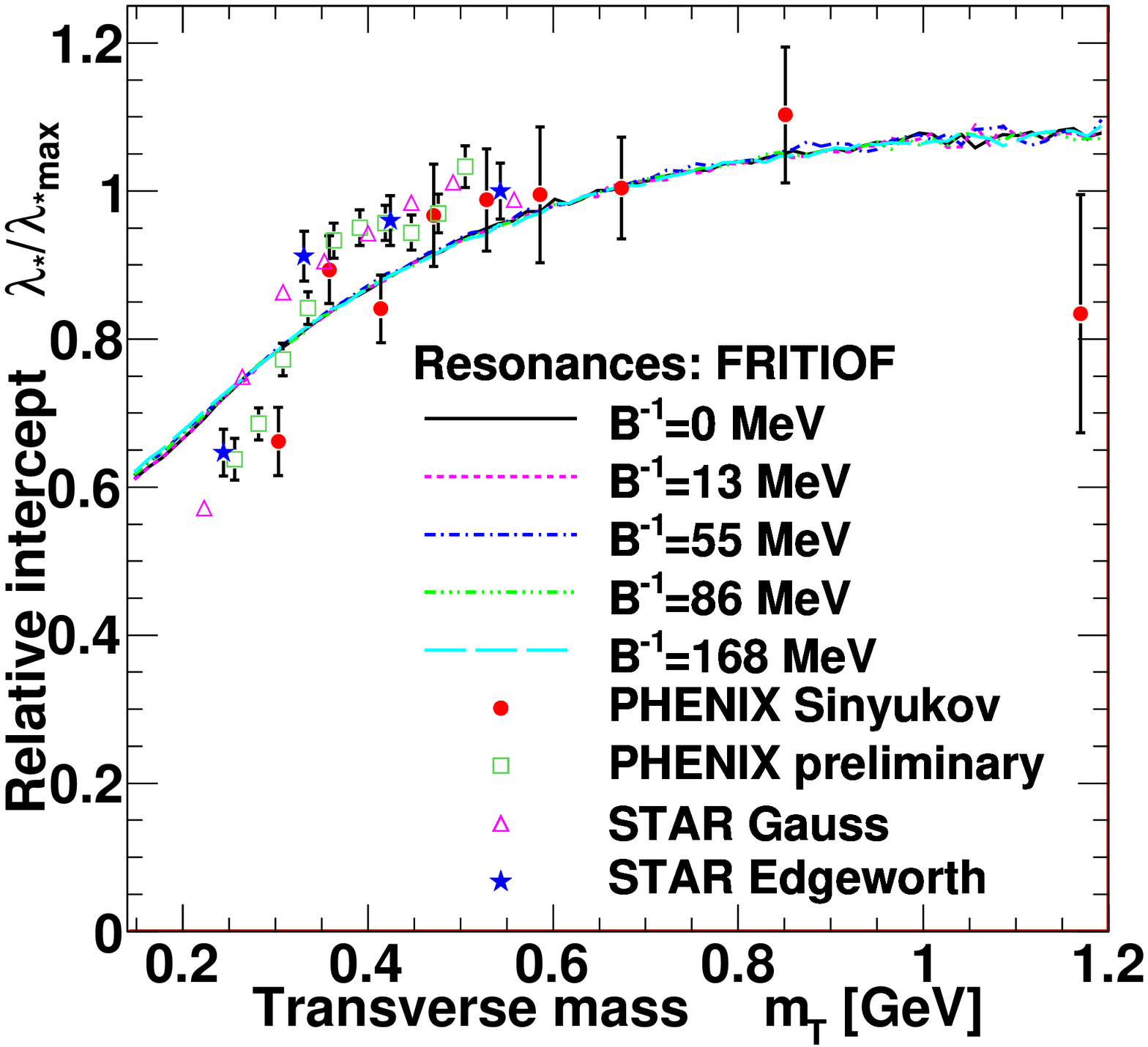}
\end{center}
\caption{{\it (Color online)}
Results for the combined STAR+PHENIX data set with resonance multiplicities from the FRITIOF model~\cite{frirqmd}. Explanation of the panels and other parameters are the same as in Fig.~\ref{fig:alcor}. Note that no values of $\Binv$ and $\meps$ provided an acceptable fit, resulting in an empty confidence level plot. This model was therefore excluded from further studies.
}
\label{fig:fritiof}
\end{figure}

\begin{figure}[h!tbp]
\begin{center}
\includegraphics[width=.50\linewidth]{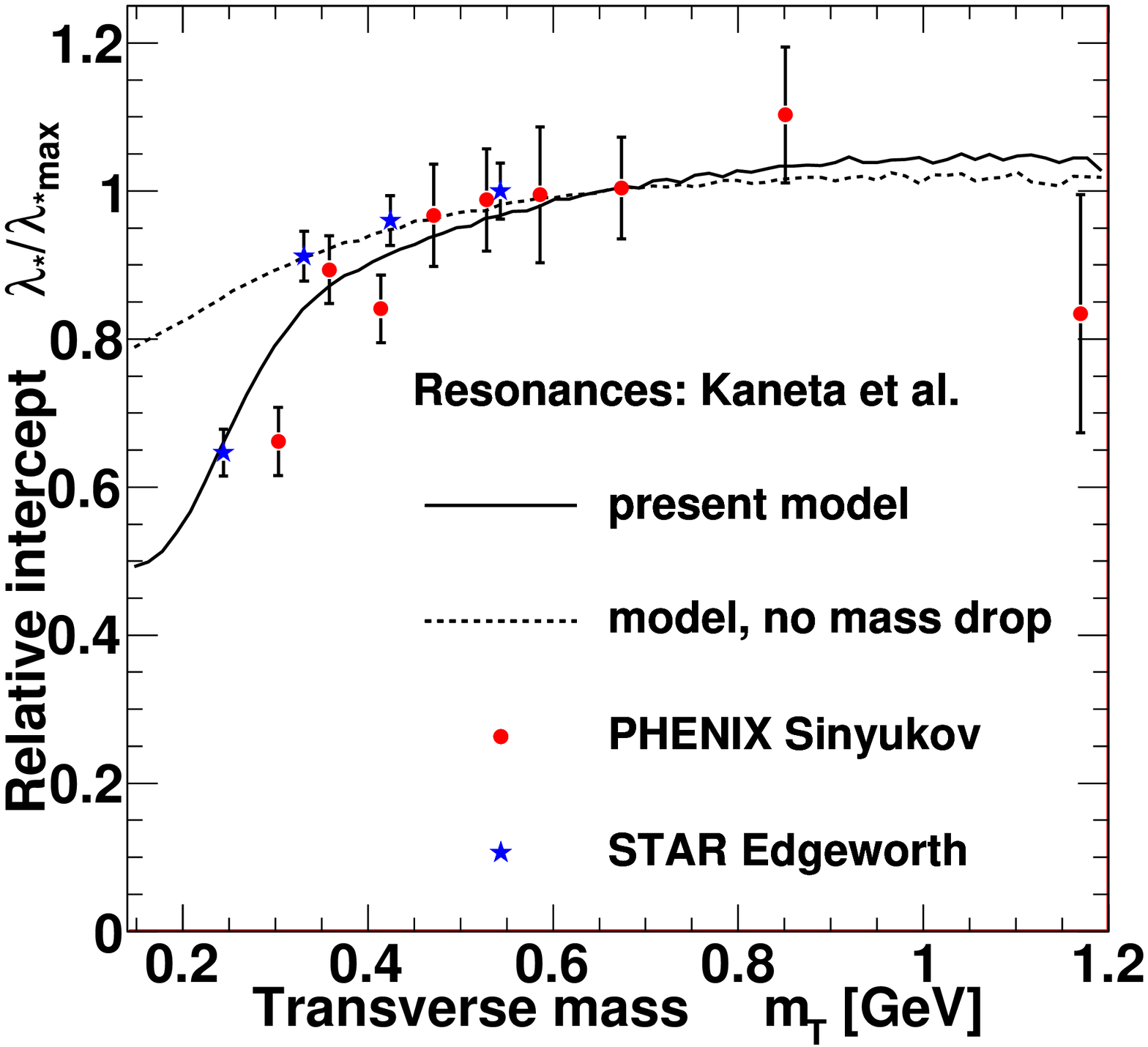}%
\includegraphics[width=.50\linewidth]{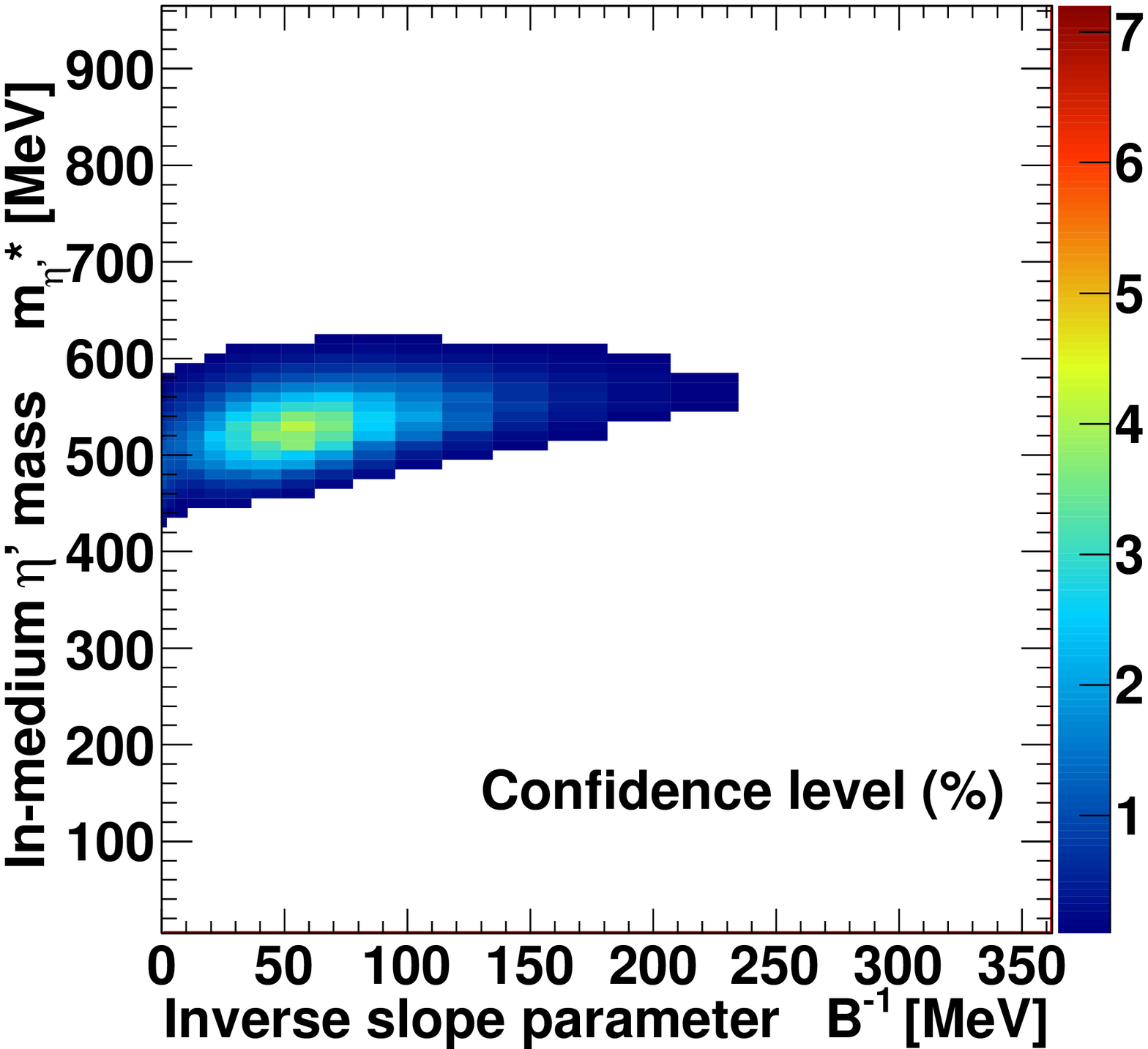}
\includegraphics[width=.50\linewidth]{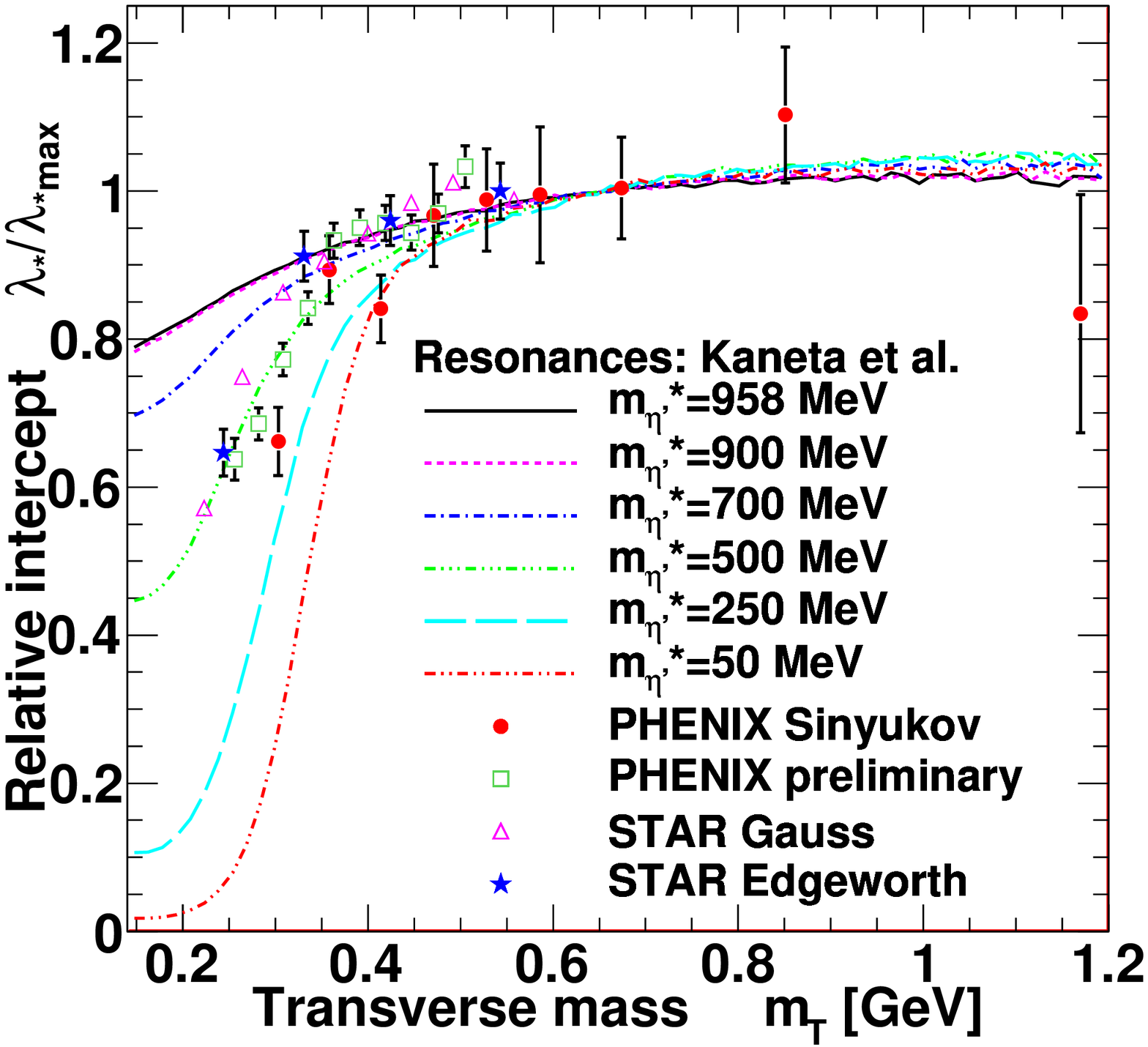}%
\includegraphics[width=.50\linewidth]{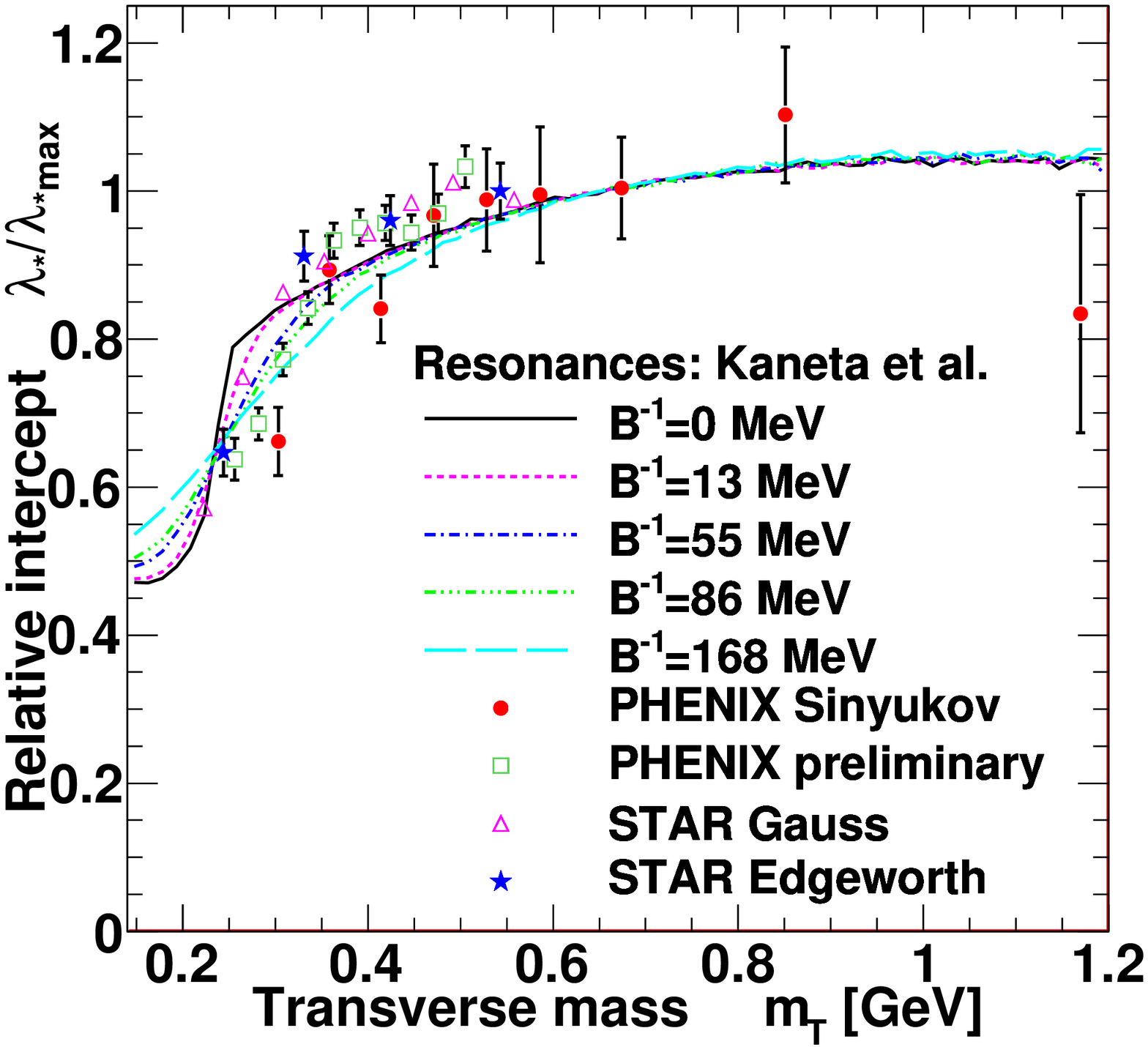}
\end{center}
\caption{{\it (Color online)}
Results for the combined STAR+PHENIX data set with resonance multiplicities from the model of Kaneta {\it et al.} \cite{kaneta}. Explanation of the panels and other parameters are the same as in Fig.~\ref{fig:alcor}. Best fit is at $\Binv=55$ MeV, $\meps=530$ MeV.
}
\label{fig:kaneta}
\end{figure}

\begin{figure}[h!tbp]
\begin{center}
\includegraphics[width=.50\linewidth]{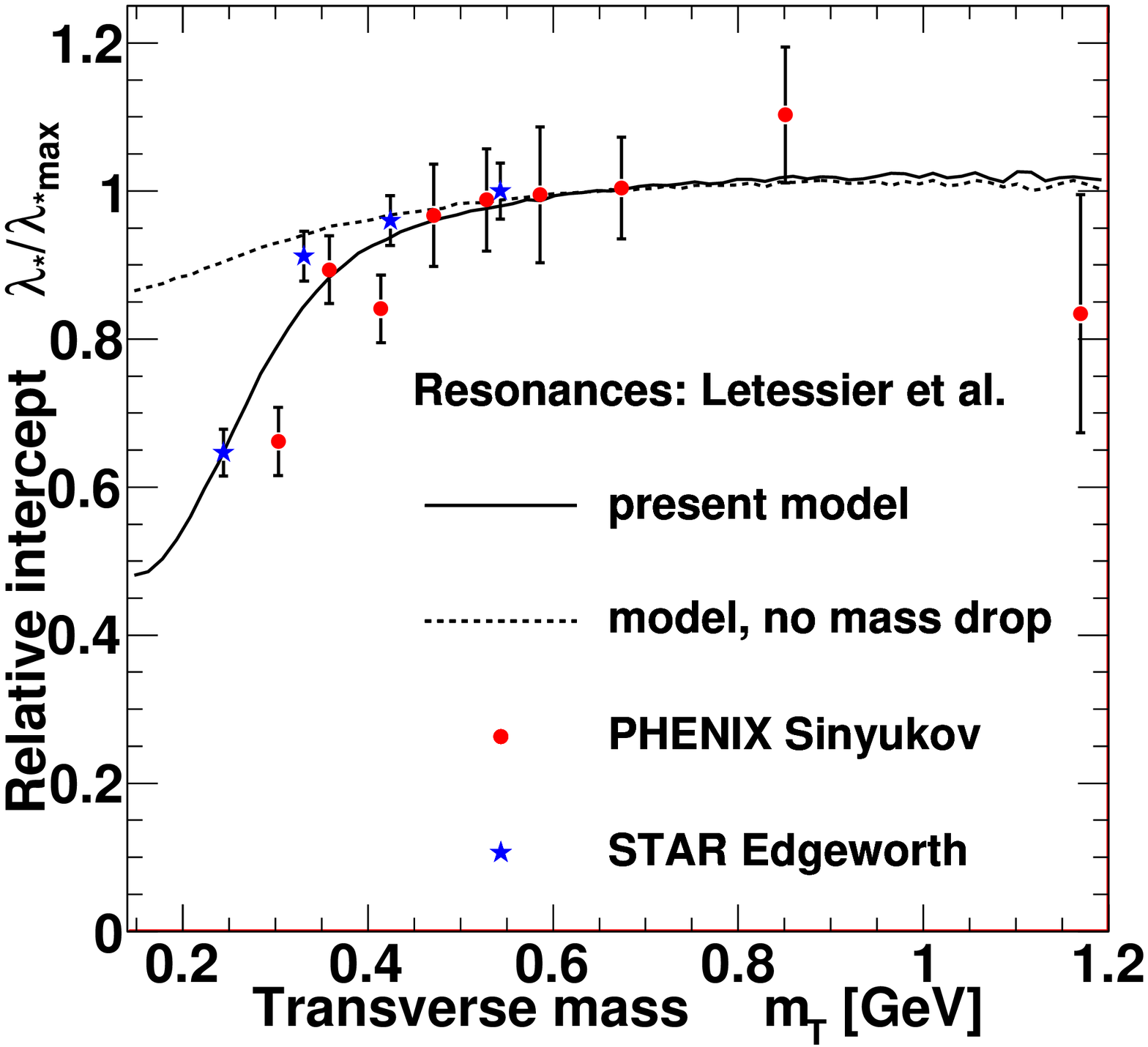}%
\includegraphics[width=.50\linewidth]{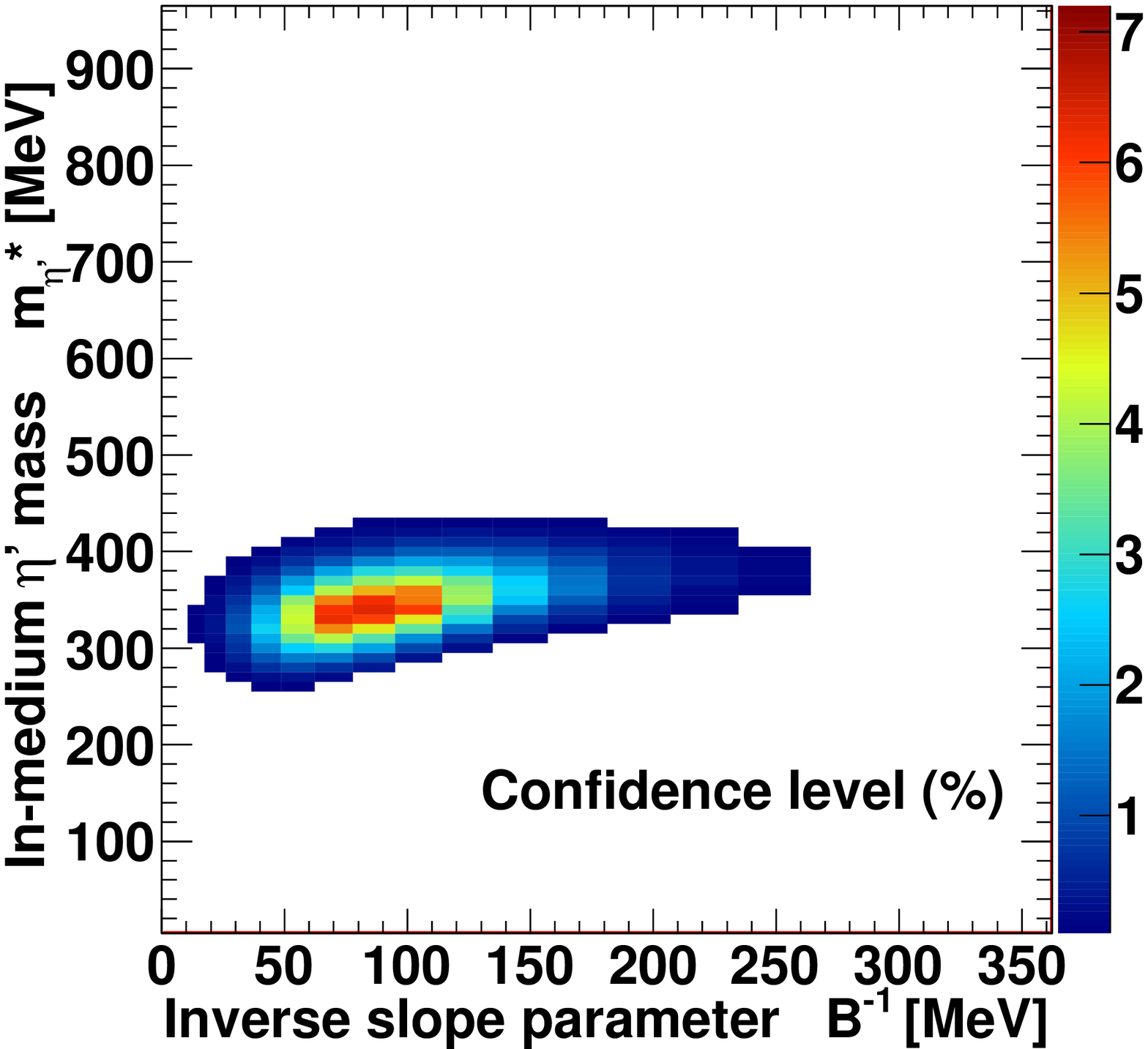}
\includegraphics[width=.50\linewidth]{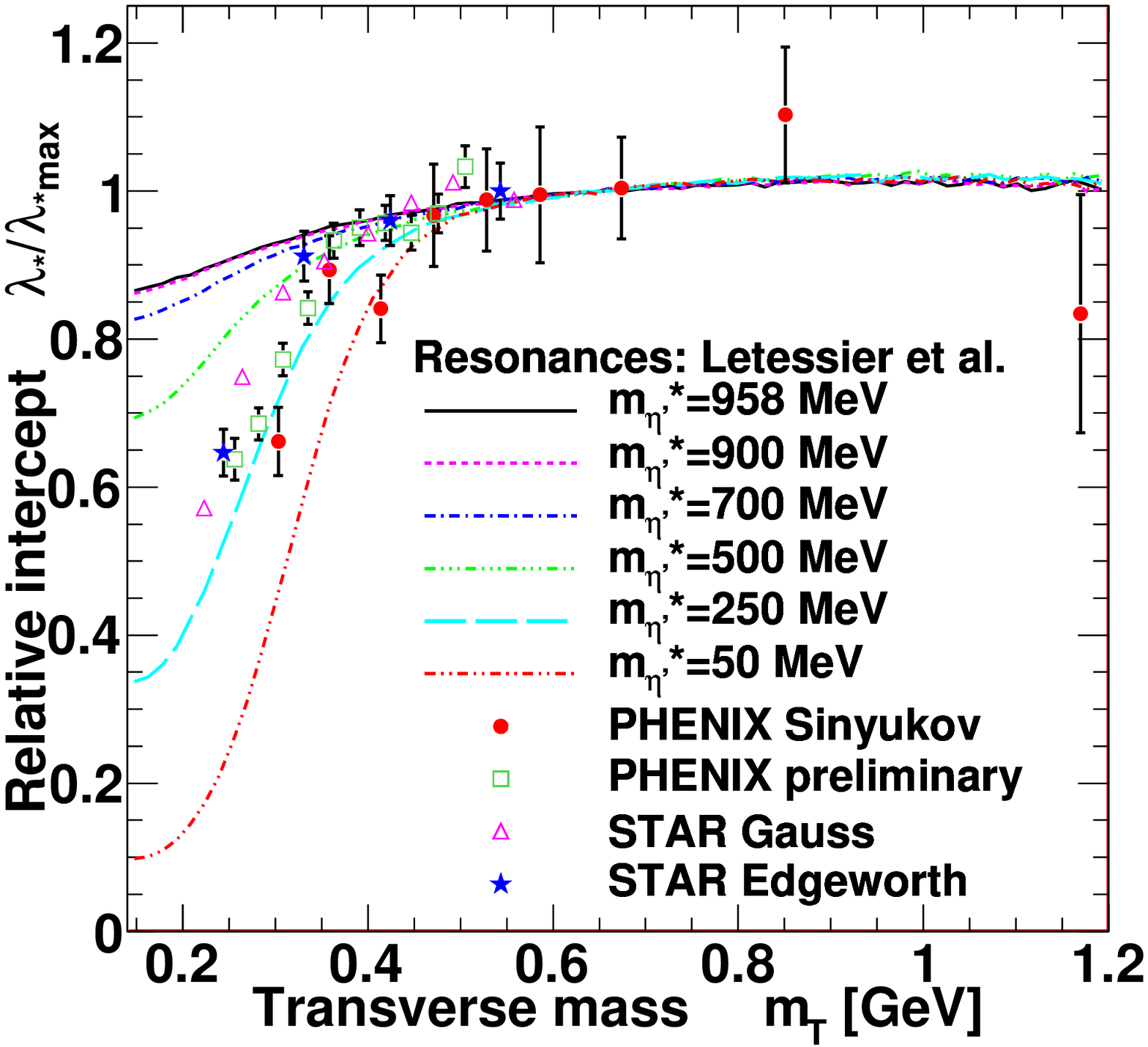}%
\includegraphics[width=.50\linewidth]{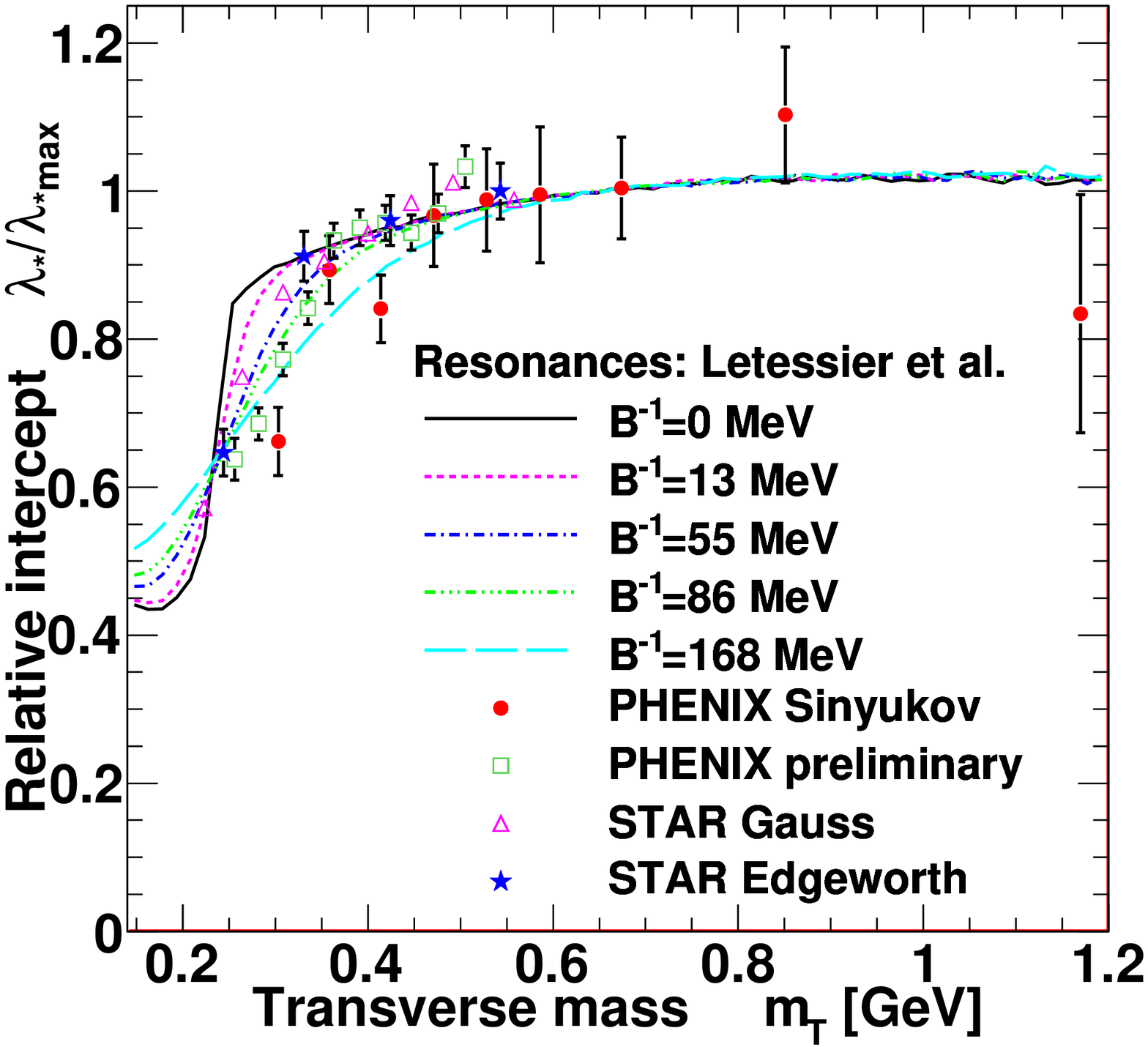}
\end{center}
\caption{{\it (Color online)}
Results for the combined STAR+PHENIX data set with resonance multiplicities from the model of Letessier {\it et al.}~\cite{rafelski}. Explanation of the panels and other parameters are the same as in Fig.~\ref{fig:alcor}. Best fit is at $\Binv=86$ MeV, $\meps=340$ MeV.
}
\label{fig:rafelski}
\end{figure}

\begin{figure}[h!tbp]
\begin{center}
\includegraphics[width=.50\linewidth]{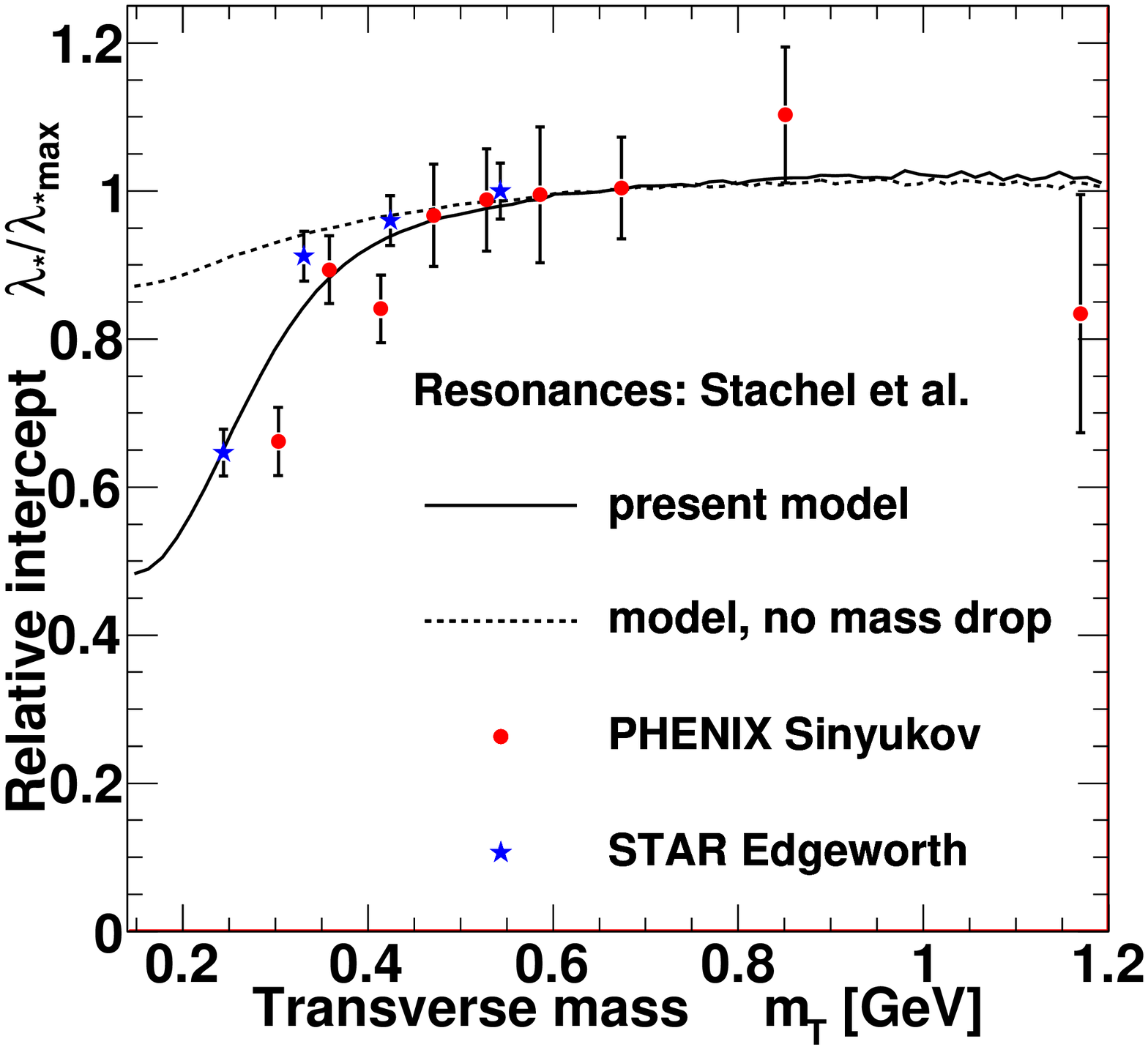}%
\includegraphics[width=.50\linewidth]{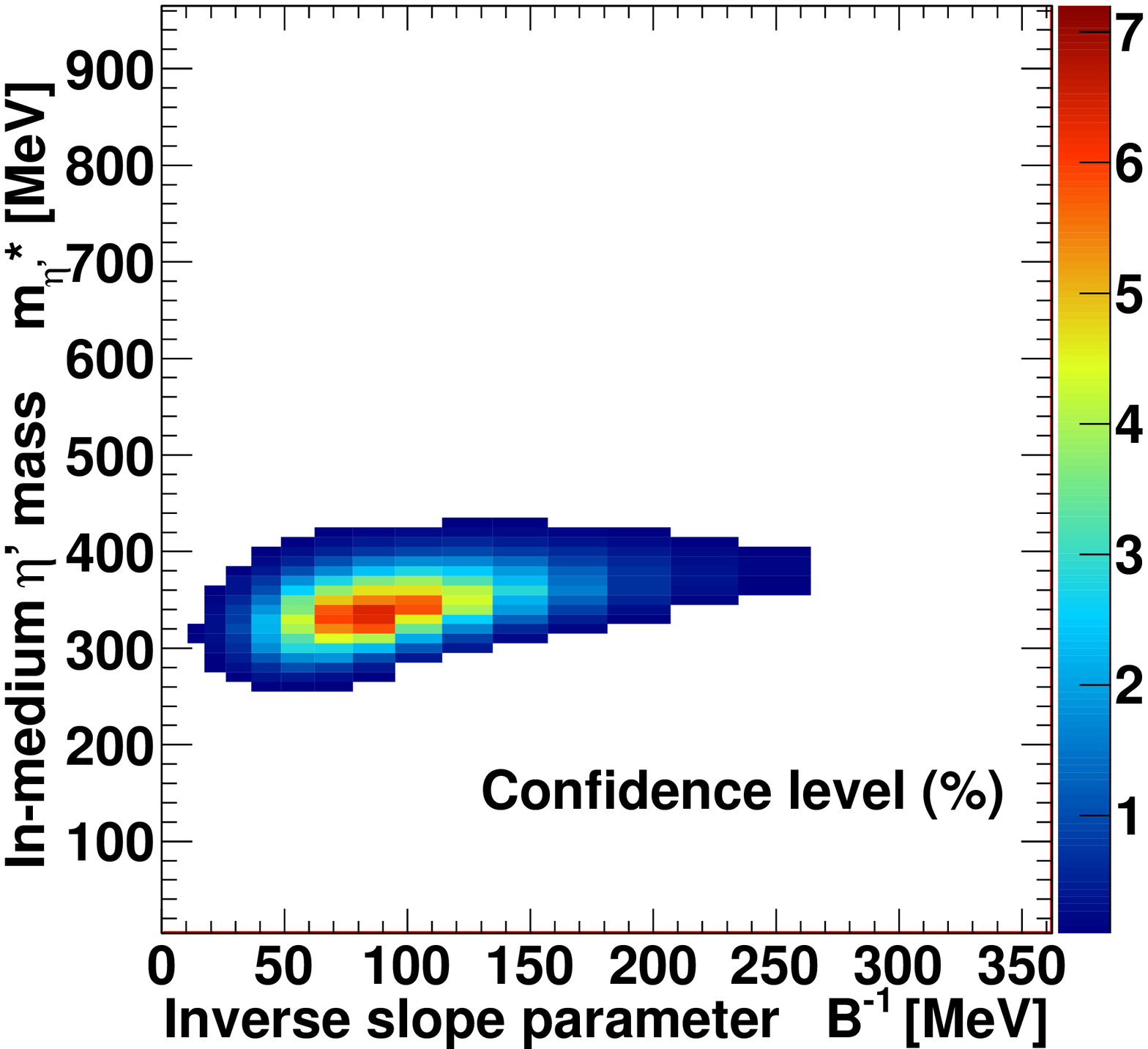}
\includegraphics[width=.50\linewidth]{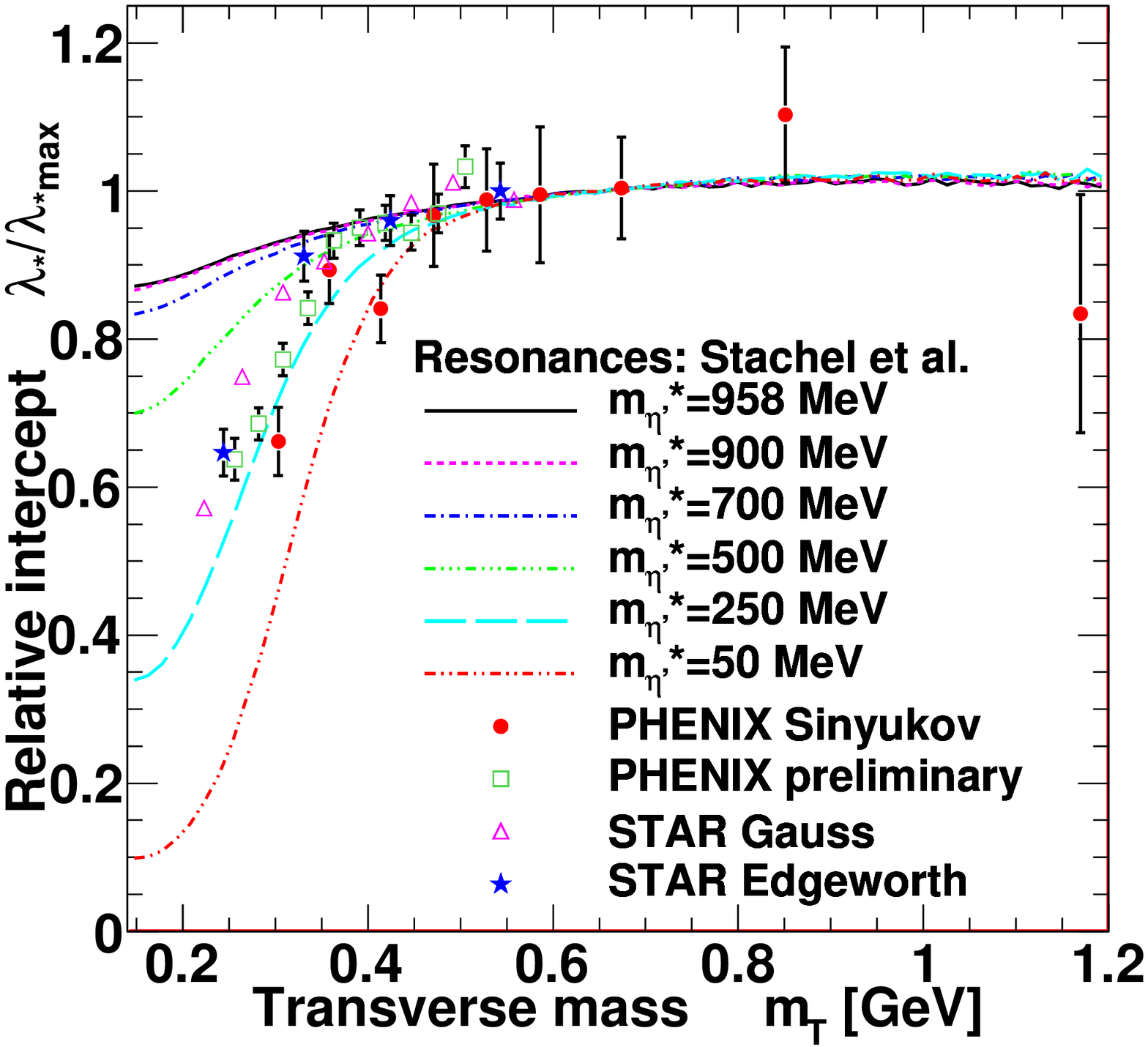}%
\includegraphics[width=.50\linewidth]{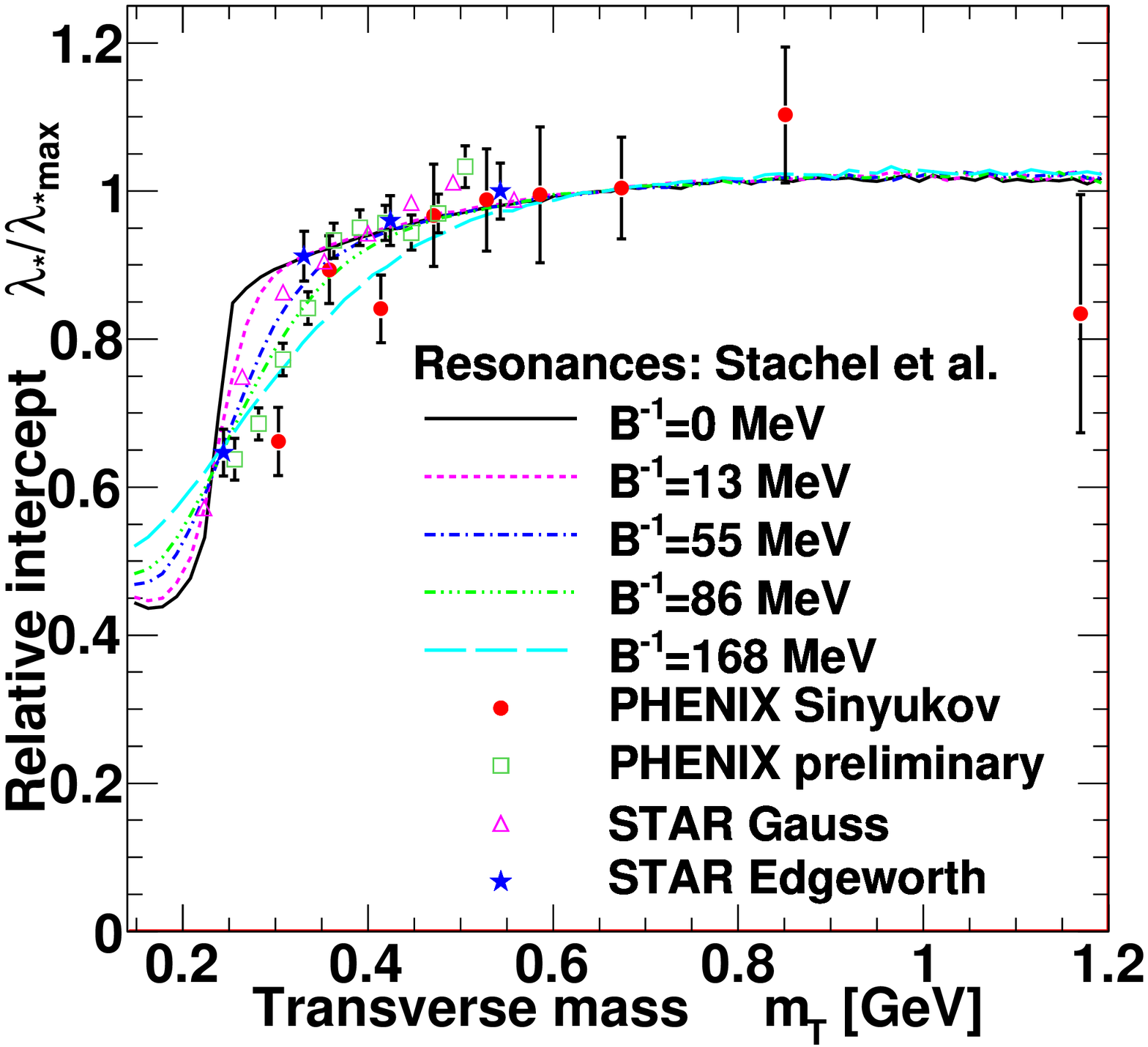}
\end{center}
\caption{{\it (Color online)}
Results for the combined STAR+PHENIX data set with resonance multiplicities from the model of  Stachel {\it et al.} \cite{stachel}. Explanation of the panels and other parameters are the same as in Fig.~\ref{fig:alcor}. Best fit is at $\Binv=86$ MeV, $\meps=340$ MeV.
}
\label{fig:stachel}
\end{figure}

\begin{figure}[h!tbp]
\begin{center}
\includegraphics[width=.50\linewidth]{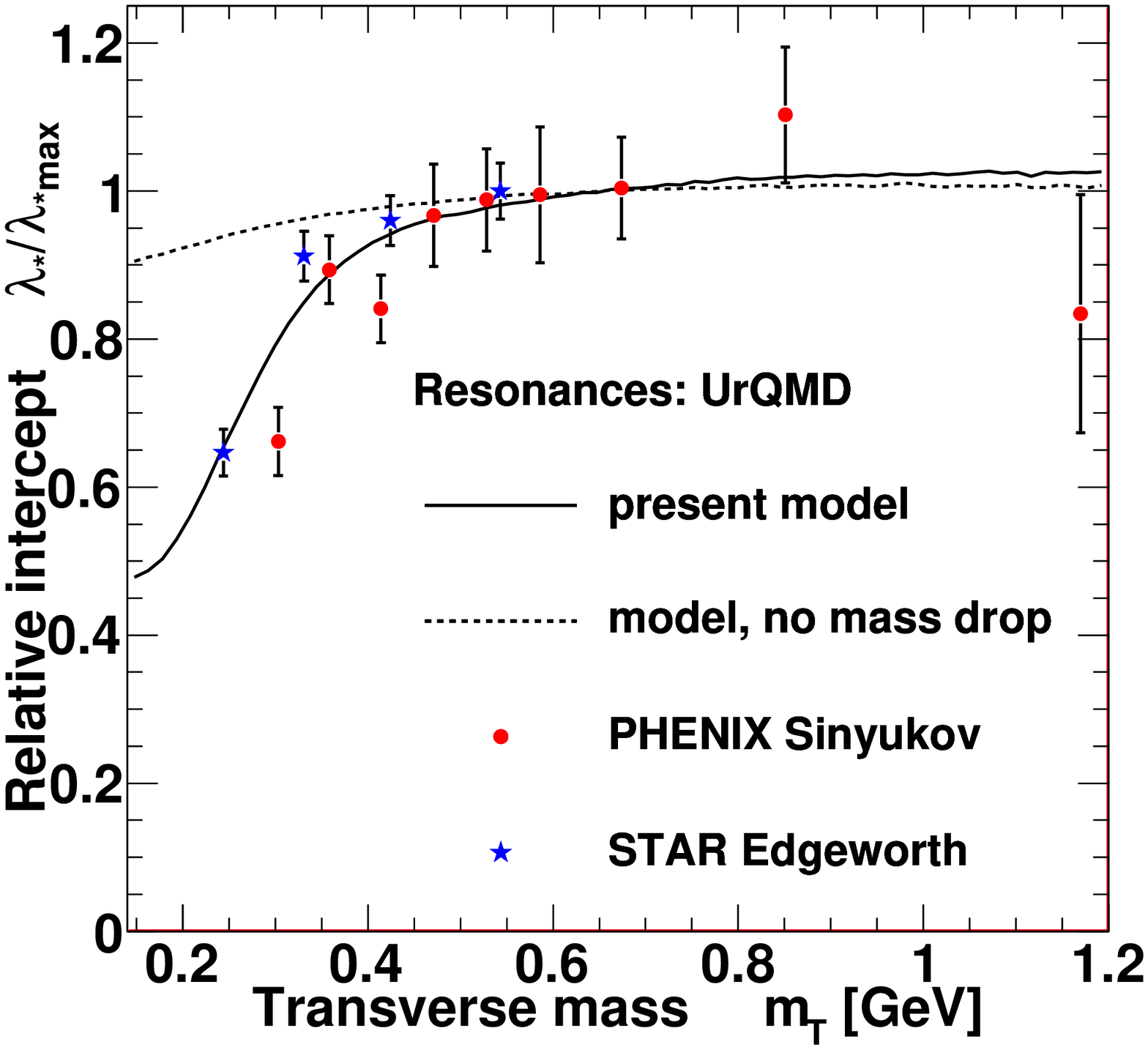}%
\includegraphics[width=.50\linewidth]{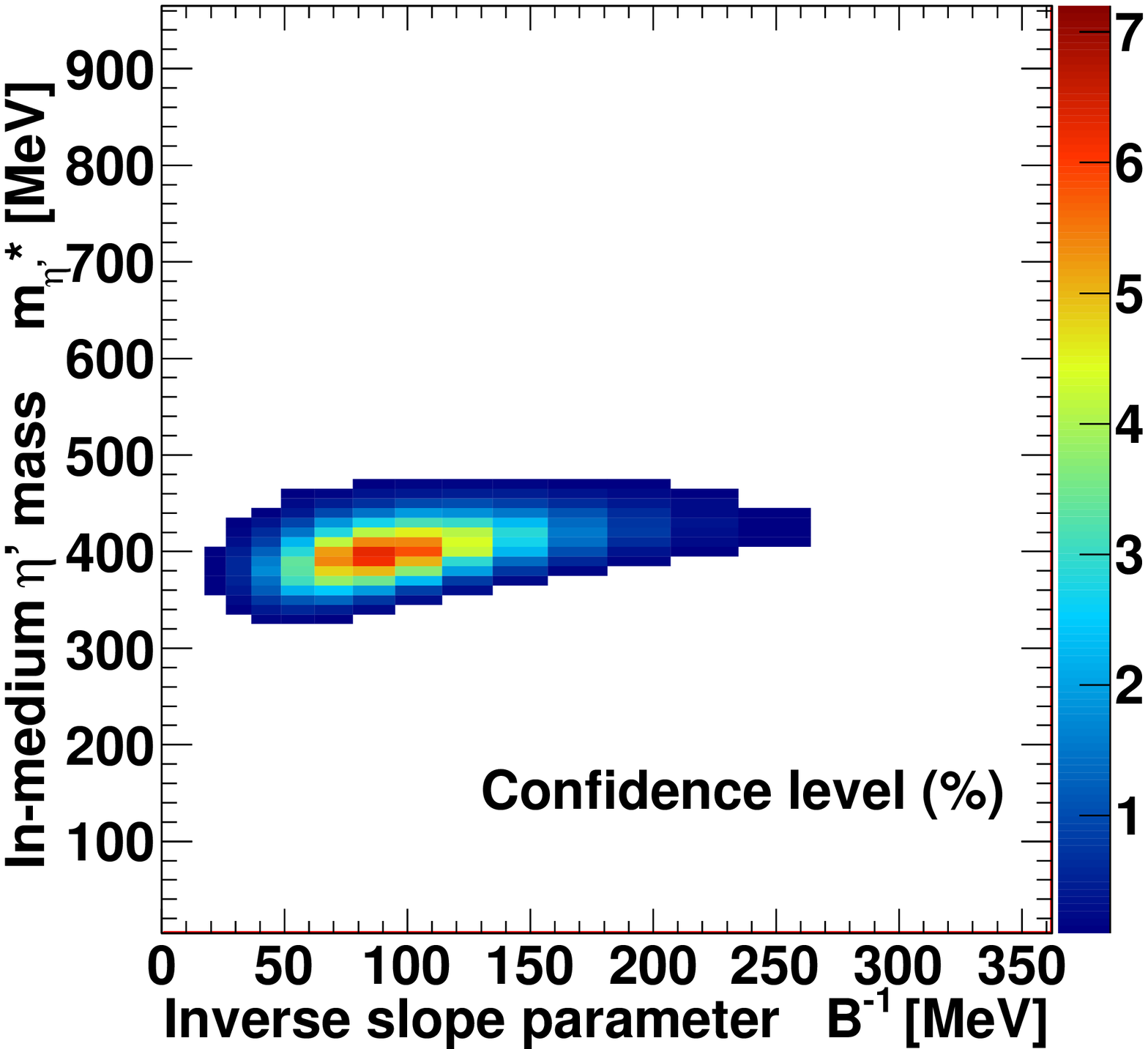}
\includegraphics[width=.50\linewidth]{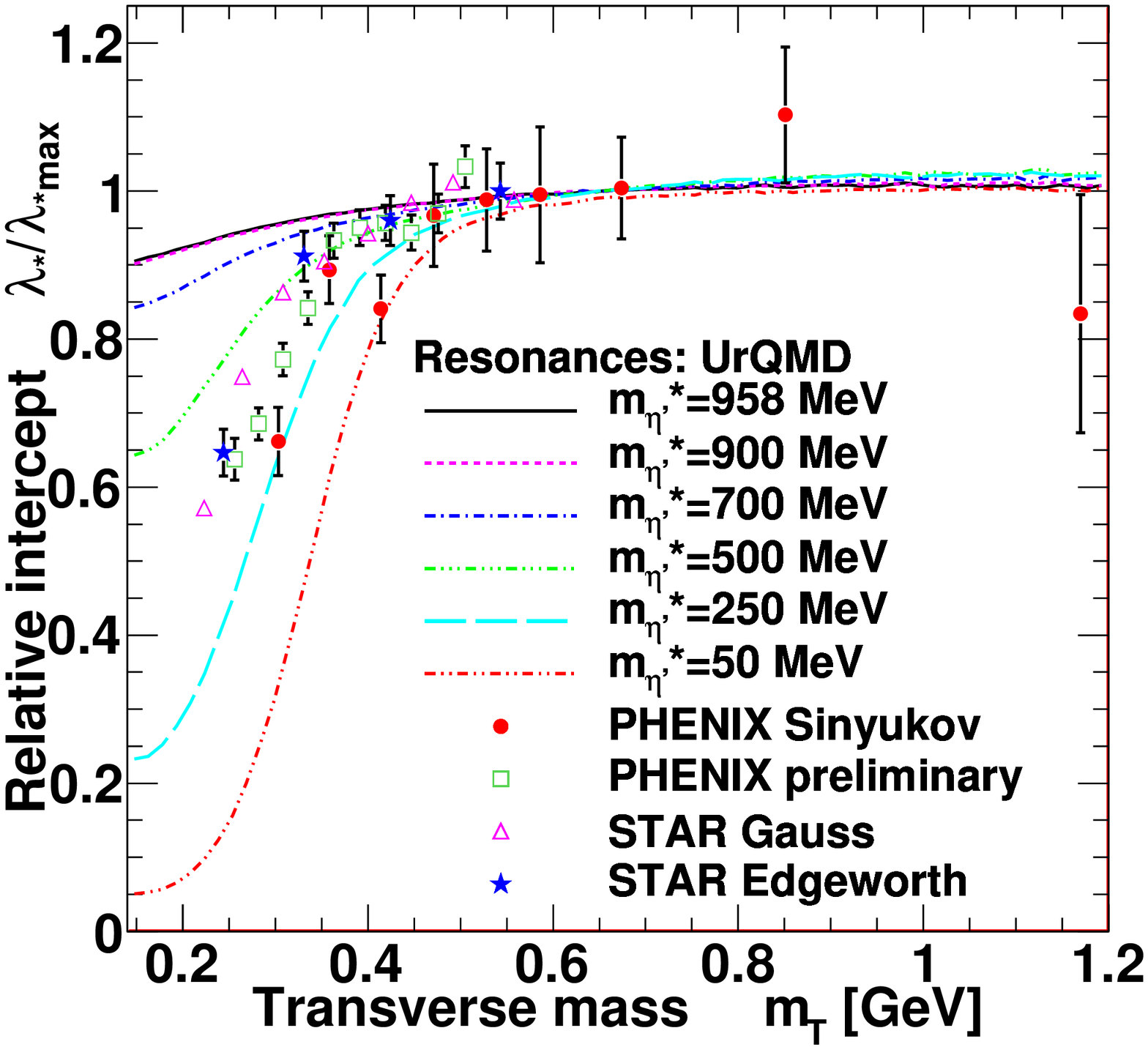}%
\includegraphics[width=.50\linewidth]{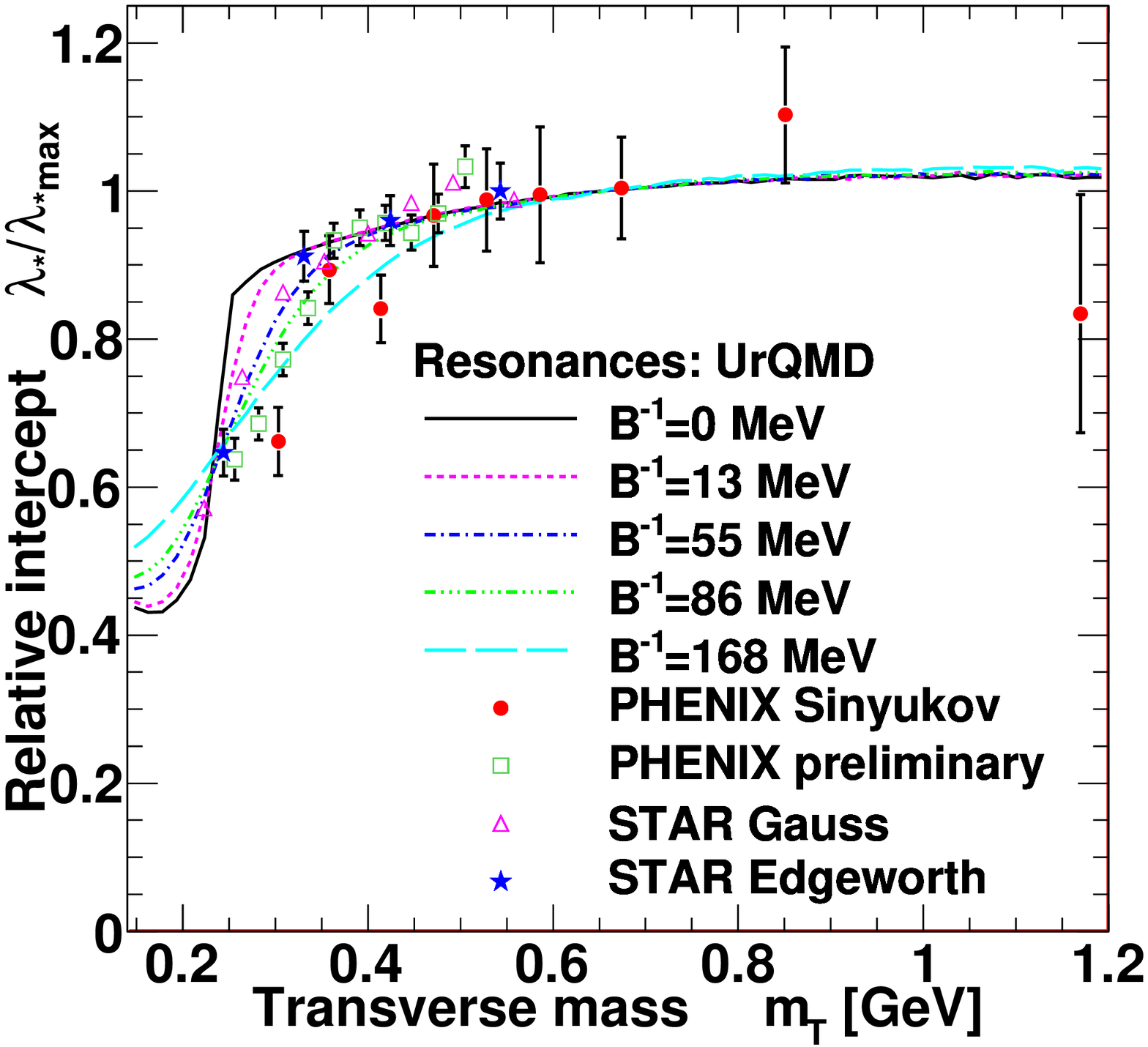}
\end{center}
\caption{{\it (Color online)}
Results for the combined STAR+PHENIX data set with resonance multiplicities from the UrQMD model~\cite{urqmd}. Explanation of the panels and other parameters are the same as in Fig.~\ref{fig:alcor}. Best fit is at $\Binv=86$ MeV, $\meps=400$ MeV.
}
\label{fig:urqmd}
\end{figure}

\begin{figure}[h!tbp]
\begin{center}
\includegraphics[width=.6\linewidth]{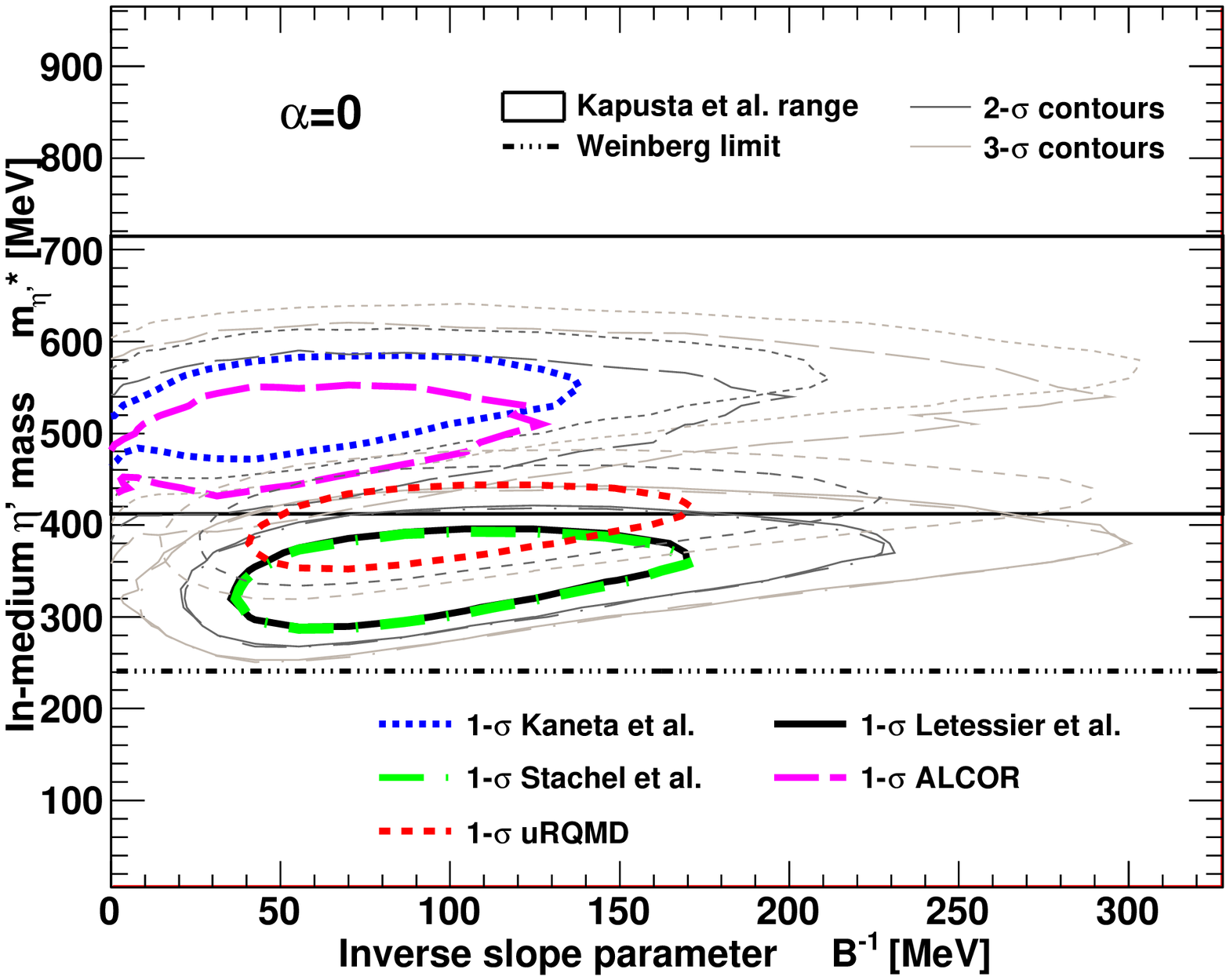}
\end{center}
\caption{{\it (Color online)}
  Standard deviation contours on the (\Binv{}, \metap) plain, obtained from 
  \lamfrac\ of MC simulations for $\alpha=0$, $\Tfo=\Tcond=177$ MeV and $\uT=0.48$, based on different chemical freeze-out models, each fitted to the PHENIX and STAR combined data set. The framed band indicates
  the theoretically predicted range of $412\ \MeV$--$715\ \MeV$~\cite{kapusta}, while the horizontal dashed-dotted line at $241\ \MeV$ indicates Weinberg's lower limit~\cite{Weinberg:1975ui}.
}
\label{fig:map_cl}
\end{figure}

\begin{figure}[h!tbp]
\begin{center}
\includegraphics[width=.6\linewidth]{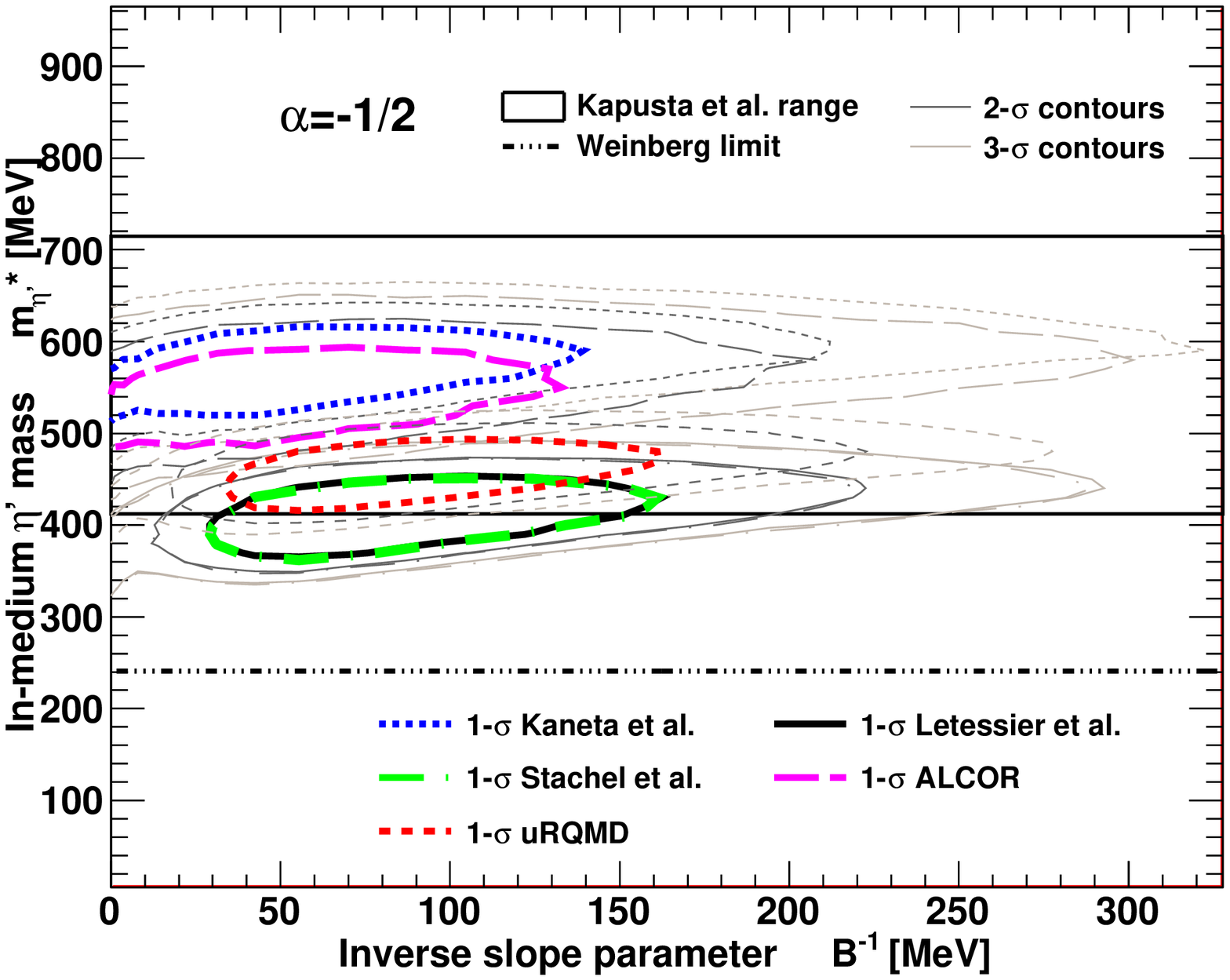}
\end{center}
\caption{{\it (Color online)}
  Standard deviation contours on the (\Binv{}, \metap) plain, $\alpha=-1/2$. 
  All the other settings are the same as in Fig.~\ref{fig:map_cl}.
}
\label{fig:map_clA}
\end{figure}

The (\Binv, \mT) plane was mapped on a grid of 21 non-equidistant steps\footnote{In fact, $B^{-1/2}$ 
was mapped in 21 equidistant steps between 0 and 600 $\MeV^{1/2}$.} for $0\ \MeV\le\Binv\le 350\ \MeV$, and by 10 MeV equidistant steps\footnote{Thus all the \meps values are given with the resolution of $10\ \MeV$ throughout the manuscript.} for the $10\ \MeV\le\metap\le 958\ \MeV$ region.
A good simultaneous description of PHENIX and STAR data was obtained by certain values of \Binv and \mT in the case of five out of the six models, as demonstrated in Figures~\ref{fig:alcor}--\ref{fig:urqmd}.
For each model, we show the confidence level ($CL$) map, the best result, the result without the mass modification, and the effect of \meps and \Binv scans around the best result. 
The standard deviation contours on (\Binv{}, \metap) plain for the best fits of \lamfrac\ simulations to data are summarized for different models in Figs.~\ref{fig:map_cl}, and \ref{fig:map_clA} for $\alpha=0$ and $\alpha=-1/2$ respectively.

%
%
%
%
%
%

%

As expected, a strongly reduced $\etap$ mass results in a dip in the low-\mT{} part of the $\lambdas(\mT)$ 
distribution (and the smaller the mass, the deeper the dip is), while $\Binv$ steers the width of this dip 
through smearing the momentum, while the $\uT$ parameter influences the overall slope of the $\lambdas(\mT)$ 
spectrum at higher \mT. Such a role of $\uT$ was already observed in~\cite{vance}, and it is not detailed here.

The systematic uncertainties resulting from the lack of precise knowledge of model constants were 
mapped by varying these constants within the range that is given by different measurements:
Spectra with values of $\alpha=-1/2$ and $\alpha=+1/2$ were investigated
as well as $\Tfo=100$ MeV, $\Tcond=140$ MeV and $\Tcond=220$ MeV, 
representing the boundaries of the theoretically or experimentally acceptable region of these parameters. 

Besides the choice of the model parameters, there are systematic uncertainties resulting from uncertainties on \lambdas(\mT):
\begin{itemize}
\item The STAR Edgeworth \lambdas values were extracted from data of 0--5\% centrality, compared to the 0--30\% rapidity range of the PHENIX Sinyukov data set. We estimate that this relative error does not exceed 9.8\%. See details in Appendix~\ref{sec:centrality}.
\item Although the $\omega$ may give part of its contribution to the tail at the resolution available in STAR and PHENIX, we considered it as part of the halo. This introduces a relative systematic error of 7\% on the measured \lambdas{}~\cite{Nickerson:1997js}. 
\item Moving the pseudorapidity cutoff from $|y|<0.36$ to $|y|<0.50$ adds another error, that is measured to be 3\%. 
\end{itemize}
These errors are considered independent and are added up quadratically. The relative error on \lamfrac corresponds to the same amount of relative error on $\meps$ in the worst-case 
scenario.\footnote{%
This relation was obtained from the STAR Edgeworth data set with the Kaneta {\it et al.}~\cite{kaneta} multiplicities: when the lowest \mT datapoint was moved by 10\% first up and then down, the fitted \meps value changed by the same relative error.}

The best fit values and their error determinations are summarized in Table~\ref{tab:massfits}, while 
Table~\ref{tab:massbounds} shows the limits of acceptability, defined on the region, where $CL>0.1\%$. 
The combined PHENIX and STAR data cannot be described unless a significant \etap mass modification is assumed, $\meps<680\ \MeV$. 
Although this value includes the systematics from the model uncertainties, it is also subject to the uncertainties coming from the other sources listed above.\footnote{%
  We have determined these errors at the best mass fits, and applied the same absolute value in case of the mass limits. This is a conservative method since the higher the mass value is, to the lesser extent it is influenced by the same effect.
} On the basis of model simulations, studies and systematic checks, we find that a description of the combined PHENIX and STAR data set is possible with 
$CL>0.1\%$ only if an in-medium \etap mass modification of $\Delta\meps>200\ \MeV$ is utilized.
In other words, $\Delta\meps>200$~MeV at the 99.9\% confidence level, corresponding to a more than 5-$\sigma$ effect, in the considered broad model class.

%
%
\begin{table}[h!tbp]
\begin{tabular}{|l|c|c|c|c|c||c|c|c|}
\hline
\multirow{3}{*}{Data set} 
& \multicolumn{5}{c||}{Acceptability boundaries of model fits} & 
\multicolumn{3}{c|}{Parameters} \\

& ALCOR & Kaneta & Letessier & Stachel & UrQMD
& \multirow{2}{*}{$\alpha$} & \multirow{2}{*}{\Tcond} & \multirow{2}{*}{\Tfo} \\
& \cite{alcor} & {\it et al.}~\cite{kaneta} & {\it et al.}~\cite{rafelski} & {\it et al.} ~\cite{stachel} & \cite{urqmd}
& & & \\ \hline\hline
%
%
%
\meps\ (MeV) & $490{+60\atop-50}$ & $530{+50\atop-50}$ & $340{+50\atop-60}$ & $340{+50\atop-60}$ & $400{+50\atop-40}$ 
& \multirow{4}{*}{0} & \multirow{4}{*}{177} & \multirow{4}{*}{177} \\  
\Binv\ (MeV) & 42 & 55 & 86 & 86 & 86 
& & & \\
CL (\%)     & 4.29 & 4.12 & 6.35 & 6.38 & 6.28
& & & \\
$\chi^2/NDF$ & 1.83 & 2.07 & 1.72 & 1.71 & 1.72
& & & \\
\hline
\meps\ (MeV) & $540{+50\atop-40}$ & $560{+60\atop-30}$ & $410{+40\atop-40}$ & $410{+40\atop-40}$ & $460{+40\atop-40}$
& \multirow{4}{*}{$-0.5$} & \multirow{4}{*}{177} & \multirow{4}{*}{177} \\  
\Binv\ (MeV) & 55 & 55 & 86 & 86 & 86 
& & & \\
CL (\%)     & 2.80 & 3.35 & 6.07 & 5.97 & 6.14
& & & \\
$\chi^2/NDF$ & 1.96 & 2.07 & 1.73 & 1.73 & 1.73
& & & \\
\hline
\meps\ (MeV) & \ & 470 & 210 & \ & \
& \multirow{4}{*}{0.5} & \multirow{4}{*}{177} & \multirow{4}{*}{177} \\  
\Binv\ (MeV) & \ & 55 & 86 & \ & \ 
& & & \\
CL (\%)     & \ & 4.58 & 6.54 & \ & \
& & & \\
$\chi^2/NDF$ & \ & 1.82 & 1.71 & \ & \
& & & \\
\hline
\meps (MeV) & \ & 620 & 460 & \ & \
& \multirow{4}{*}{0} & \multirow{4}{*}{140} & \multirow{4}{*}{177} \\  
\Binv\ (MeV) & \ & 42 & 69 & \ & \ 
& & & \\
CL (\%)     & \ & 2.26 & 5.86 & \ & \
& & & \\
$\chi^2/NDF$ & \ & 2.02 & 1.74 & \ & \
& & & \\
\hline
\meps (MeV) & \ & 440 & 200 & \ & \
& \multirow{4}{*}{0} & \multirow{4}{*}{220} & \multirow{4}{*}{177} \\  
\Binv\ (MeV) & \ & 69 & 104 & \ & \ 
& & & \\
CL (\%)     & \ & 5.61 & 6.33 & \ & \
& & & \\
$\chi^2/NDF$ & \ & 1.75 & 1.72 & \ & \
& & & \\
\hline
\meps (MeV) & \ & 410 & 240 & \ & \
& \multirow{4}{*}{0} & \multirow{4}{*}{177} & \multirow{4}{*}{100} \\  
\Binv\ (MeV) & \ & 145 & 145 & \ & \ 
& & & \\
CL (\%)     & \ & 1.63 & 1.80 & \ & \
& & & \\
$\chi^2/NDF$ & \ & 2.11 & 2.09 & \ & \
& & & \\
\hline
\end{tabular}
\caption{\label{tab:massfits}Fitted values of the modified \etap mass on the STAR+PHENIX combined data set, for different resonance models and parameters. The FRITIOF model has CL$<0.1\%$ and therefore it is not shown here. The statistical errors are given by the 1-$\sigma$ boundaries of the fits, determined only for \meps\ and for the $\alpha=0$ and $\alpha=-0.5$ simulations. Best \meps and \Binv parameters for various systematic checks are shown in the last four rows.
}
\end{table}
%
\begin{table}[h!tbp]
\begin{tabular}{|l|c|c|c|c|c|c||c|c|c|}
\hline
\multirow{3}{*}{Data set} 
& \multicolumn{6}{c||}{Acceptability boundaries of model fits} & 
\multicolumn{3}{c|}{Parameters} \\

& ALCOR & FRITIOF & Kaneta & Letessier & Stachel & UrQMD
& \multirow{2}{*}{$\alpha$} & \multirow{2}{*}{\Tcond} & \multirow{2}{*}{\Tfo} \\
& \cite{alcor} & \cite{frirqmd} & {\it et al.}~\cite{kaneta} & {\it et al.}~\cite{rafelski} & {\it et al.} ~\cite{stachel} & \cite{urqmd}
& & & \\ \hline\hline

PHENIX & 0--750 & 680--958 & 0--720 & 0--510 & 0--500 & 0--530 
& \multirow{3}{*}{0} & \multirow{3}{*}{177} & \multirow{3}{*}{177} \\ 

STAR & 380--600 & none & 430--630 & 190--450 & 190--450 & 260--500 
& & & \\

combined & 380--590 & none & 420--620 & 260--430 & 260--430 & 330--470 
& & & \\
\hline

PHENIX & 30--770 & 420--958 & 50--730 & 0--540 & 0--540 & 0--560 
& \multirow{3}{*}{$-0.5$} & \multirow{3}{*}{177} & \multirow{3}{*}{177} \\ 

STAR & 470--630 & none & 500--650 & 300--500 & 300--500 & 360--540 
& & & \\

combined & 450--620 & 670--760 & 490--640 & 340--480 & 340--480 & 400--510 
& & & \\
\hline

PHENIX & \ & \ & 0--690 & 0--450 & \ & \ 
& \multirow{3}{*}{0.5} & \multirow{3}{*}{177} & \multirow{3}{*}{177} \\ 

STAR & \ & \ & 320--610 & 0--390 & \ & \ 
& & & \\

combined & \ & \ & 340--590 & 0--390 & \ & \ 
& & & \\
\hline

PHENIX & \ & \ & 0--760 & 0--450 & \ & \ 
& \multirow{3}{*}{0} & \multirow{3}{*}{140} & \multirow{3}{*}{177} \\ 

STAR & \ & \ & 560--690 & 0--390 & \ & \ 
& & & \\

combined & \ & \ & 540--680 & 0--360 & \ & \ 
& & & \\
\hline

PHENIX & \ & \ & 0--680 & 0--410 & \ & \ 
& \multirow{3}{*}{0} & \multirow{3}{*}{220} & \multirow{3}{*}{177} \\ 

STAR & \ & \ & 270--580 & 0--350 & \ & \ 
& & & \\

combined & \ & \ & 290--560 & 100--320 & \ & \ 
& & & \\
\hline

PHENIX & \ & \ & 220--470 & 30--310 & \ & \ 
& \multirow{3}{*}{0} & \multirow{3}{*}{177} & \multirow{3}{*}{100} \\ 

STAR & \ & \ & 360--470 & 190--300 & \ & \ 
& & & \\

combined & \ & \ & 370--440 & 200--280 & \ & \ 
& & & \\
\hline

\end{tabular}
\caption{\label{tab:massbounds}
Acceptability boundaries of the modified \etap mass on the PHENIX, STAR, and the combined PHENIX+STAR data sets, for different resonance models and parameters. A fit is considered acceptable if CL$\geq 0.1\%$. 
FRITIOF fails completely on the STAR data set and also on the combined PHENIX+STAR data. All the other models require an $\meps{}\leq 640~\mathrm{MeV}$ excluding systematics. Acceptability boundaries for various systematic checks are shown in the last four rows.}
\end{table}

Note that the validity of our analysis relies on the correctness of published STAR and PHENIX data. These data, however, were not measured with the definite purpose to serve as a base for the search for the in-medium \etap mass modification, where particular attention has to be payed to the momentum dependence of the particle identification purity and efficiency, especially at low-\PT regime.
More detailed dedicated $\lambdas(\mT)$ measurements together with additional analysis of the dilepton and photon decay channels of \etap could help consolidate the findings reported here.

\section{The enhanced \etap and $\eta$ spectrum}

The dilepton spectrum has been measured recently in minimum bias Au+Au collisions at \sqsn = 200 GeV,
and a large enhancement was observed in the low-invariant mass region $m_{\rm ee}<1\ \GeV$~\cite{Adare:2009qk}.
Low transverse mass enhancement of the \etap and $ \eta$ production results in dilepton enhancement just in the considered kinematic range~\cite{kapusta}. 
PHENIX recently reported a two-component transverse momentum spectrum in dilepton channel direct photon measurements~\cite{Adare:2009qk}. The \etap and $\eta$ spectra reconstructed here may serve as an input, e.g., for the simulations and evaluations of the dilepton spectra instead of the currently utilized $\etap$ spectra, based on $\mT$ scaling and hence not providing the possibility of taking into account an \etap mass reduction. 

In order to make a cross-check possible with measurements that are based on the non-pionic decay channels of the $\etap$, we extract the transverse momentum  
spectra of the $\etap$ and $\eta$ mesons from the Bose--Einstein correlation measurements at mid-rapidity in the reduced \meps scenario, for each of the successful resonance models.

\begin{figure}[h!tbp]
\begin{center}
\includegraphics[width=.5\linewidth]{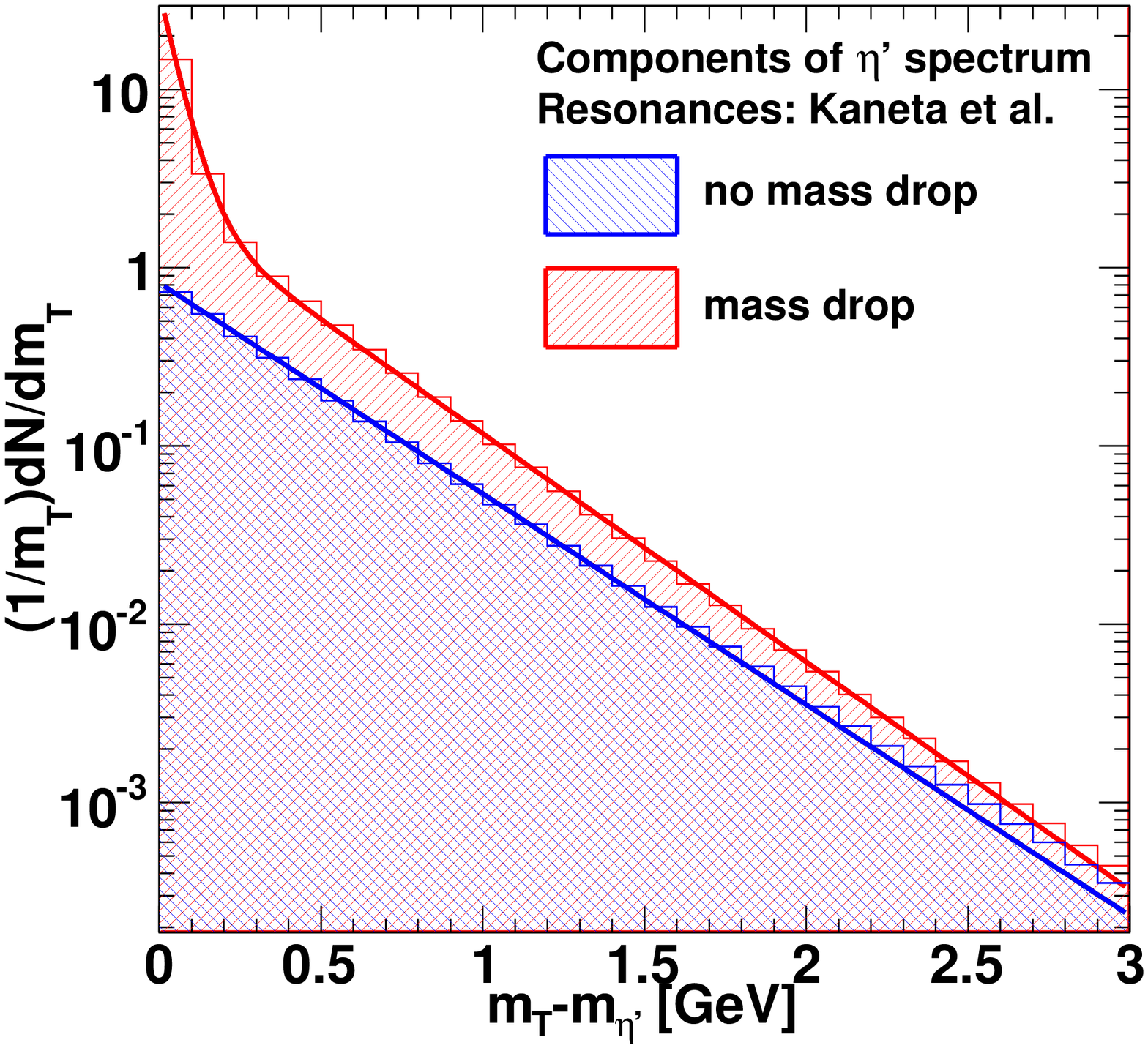}%
\includegraphics[width=.5\linewidth]{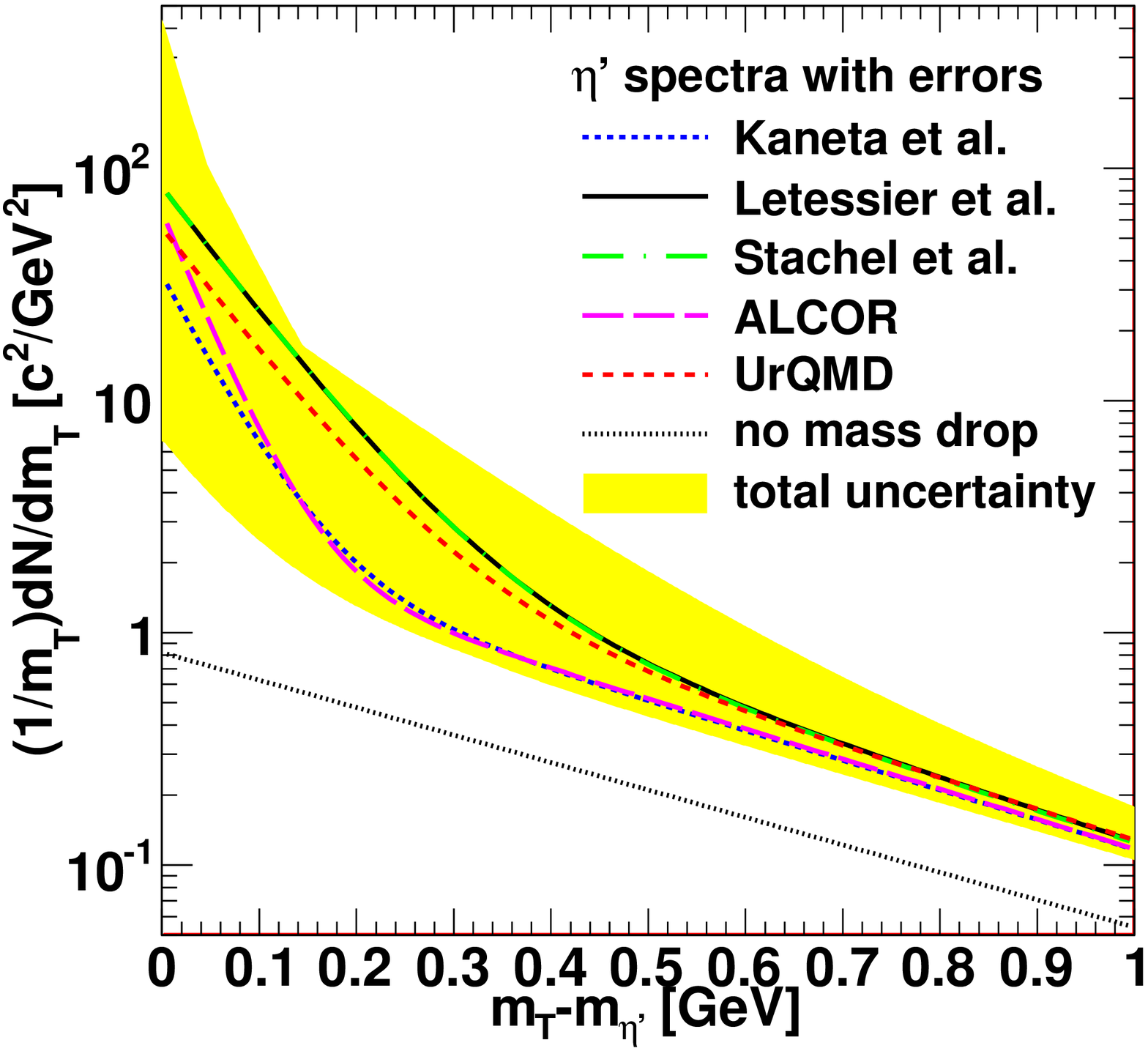}
\end{center}
\caption{{\it (Color online)}
{\it Left:} Reconstructed $\mT$ spectrum of the $\eta^\prime$ mesons using the resonance ratios of Ref.~\cite{kaneta}.
Lower (blue) part indicates the scenario without in-medium $\eta^\prime$ mass reduction, 
upper (red) part the enhancement required to describe
the dip in the low $\mT$ region of $\lambdas$.
{\it Right:} 
Comparison of reconstructed $\mT$ spectra of the $\eta^\prime$ mesons from different models. 
All spectra are normalized to the \etap multiplicity as given by the model of Kaneta and Xu~\cite{kaneta}, and then fitted with a double exponential. The shaded (yellow) band represents the total error. The fit parameters are summarized in Table~\ref{tab:spectra}.
Above $\mT-\metap = 1\ \GeV$, all models result in very similar values, corresponding to an approximate \mT{}-scaling. This figure indicates that the violation of this \mT{}-scaling is model dependent, and suggests 
that dilepton measurements may provide additional constraints for the model builders. 
}
\label{fig:spec}
\end{figure}

We compute the spectra two times for each model. First we use the original $\metap=958\ \MeV$ value. The obtained spectra clearly show the $\mT$ scaling. Then we use the reduced \etap mass, and the corresponding \Binv value that provides the best description of the data in the frame of a given resonance model. These spectra break the $\mT$ scaling with an additional, steeper exponential-like part over the original \etap spectrum, and produce most of the \etap enhancement in the low-\PT region. The original ``non-enhanced'' spectra is normalized to the \etap multiplicity\footnote{
The normalization of the original spectra is performed the following way: First the numerical integral of the unmodified 
$\frac{1}{\mT}\frac{dN}{d\mT}$ distribution is computed. Then the average number of \etap mesons per event is divided by this integral, and the histogram is scaled by the resulting number.
}
predicted by Kaneta {\it et al.}~\cite{kaneta}, and then the enhanced spectra is scaled relative to it. Left panel of Fig.~\ref{fig:spec} shows both the original, \mT-scaling spectra, and the enhanced version with $\meps<m_\etap$ for the resonance ratios of Ref.~\cite{kaneta}. A comparison of enhanced \etap spectra of all resonance models is given on the right panel of Fig.~\ref{fig:spec}, and the fitted spectrum parameters are listed in Table~\ref{tab:spectra} together with the \etap enhancement factors for each particular model of resonance ratios.

%
\begin{table}[h!tbp]
\begin{tabular}{|c|c|c|c|c|c|}
\hline Resonance model & $a$ & $b$ & $c$ & $d$ & $f_\etap$ \\
\hline
\hline
No enhancement
& 0.82 & 2.72 & 0 (fixed)& 0 (fixed) & - \\
\hline
ALCOR~\cite{alcor}
& 2.30 & 2.98 & 62.4 & 23.5 & 43.4 \\
Kaneta~{\it et al.}~\cite{kaneta}
& 2.21 & 2.94 & 32.4 & 18.7 & 25.6 \\
Letessier~{\it et al.}~\cite{rafelski}
& 2.91 & 3.14 & 80.1 & 12.8 & 67.6 \\
Stachel~{\it et al.}~\cite{stachel}
& 2.85 & 3.13 & 80.0 & 12.8 & 67.6 \\
UrQMD~\cite{urqmd}
& 2.75 & 3.07 & 52.5 & 12.7 & 45.0 \\
\hline 
\end{tabular}
\caption{\label{tab:spectra}
\etap enhancement ($f_\etap$) and fit parameters of the spectra for different models of resonance abundancies. The spectra are obtained using the most probable \Binv, \meps parameters, and then fitted with a double exponential function
$y = a \mathrm{e}^{-b(\mT-\metap)} + c \mathrm{e}^{-d(\mT-\metap)}$. 
The spectrum without enhancement is computed with the Kaneta {\it et al.}~\cite{kaneta} resonance ratios, with \meps=\metap.
}
\end{table}

\begin{figure}[h!tbp]
\begin{center}
\includegraphics[width=.5\linewidth]{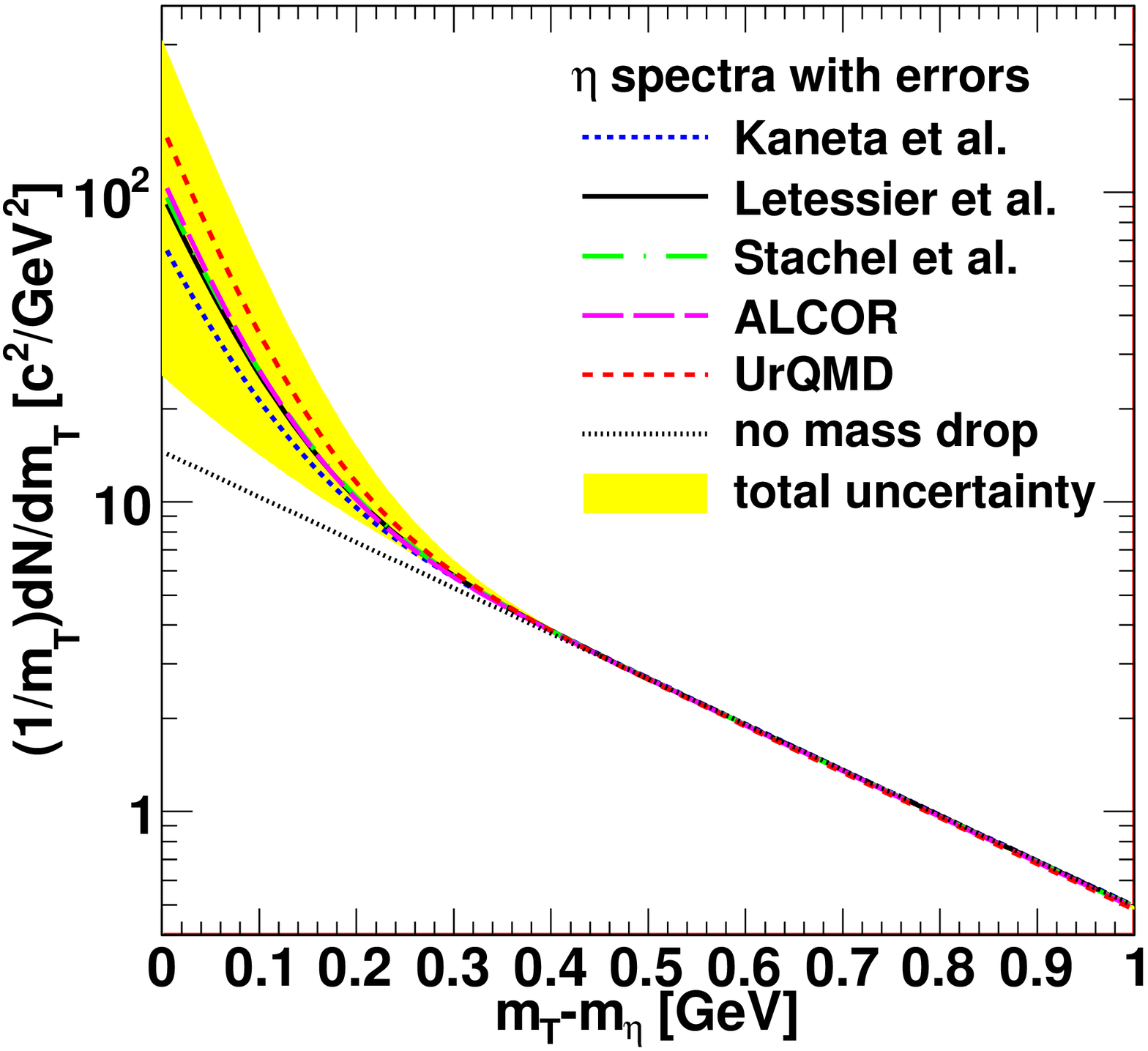}%
\includegraphics[width=.5\linewidth]{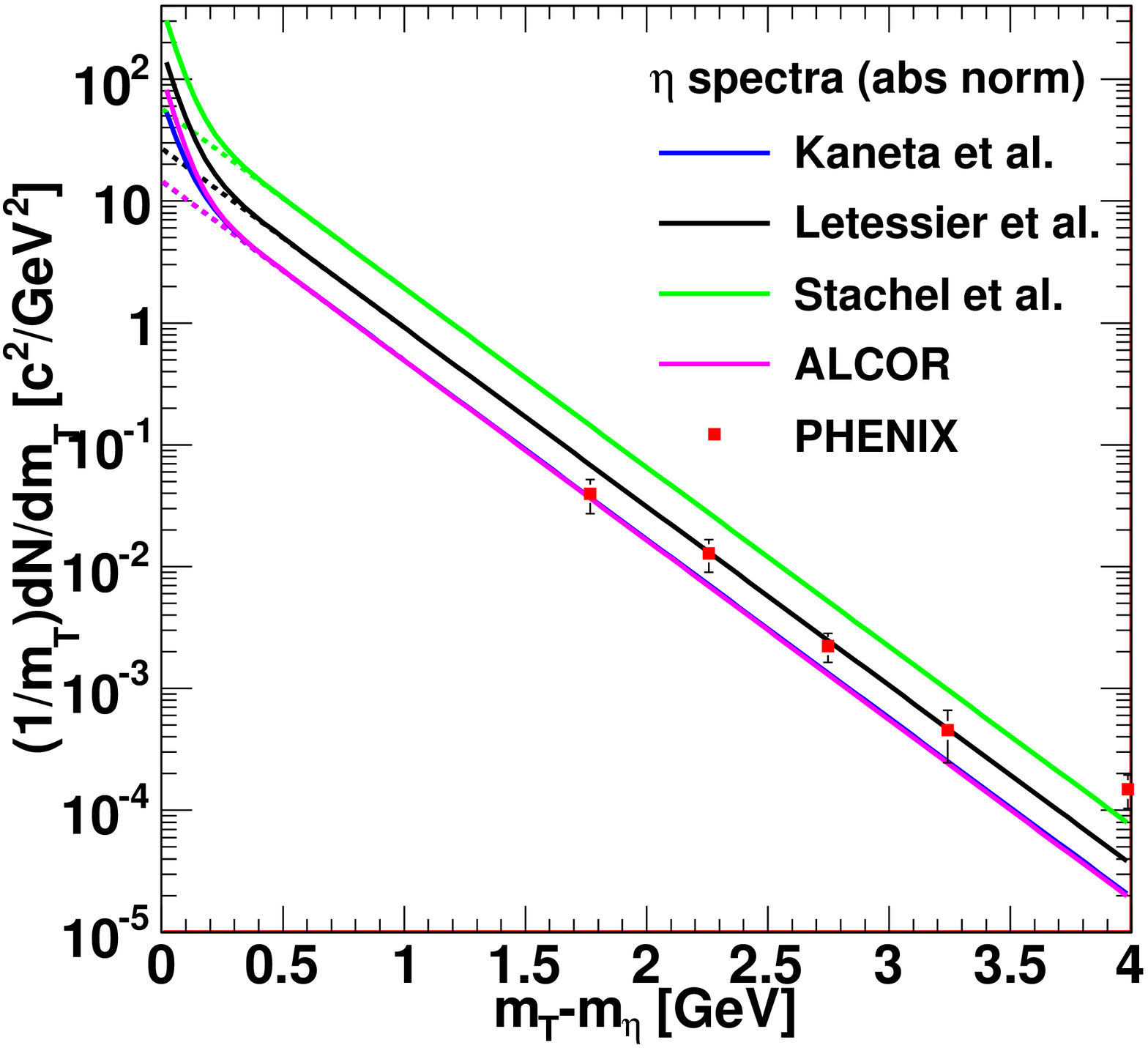}
\end{center}
\caption{{\it (Color online)}
Comparison of reconstructed $\mT$ spectra of the $\eta$ mesons from different models. 
{\it Left:} All spectra are normalized to the \etap multiplicity as given by the model of Kaneta and Xu~\cite{kaneta}, and then fitted with a double exponential. 
The shaded (yellow) band represents the total error.
The fit parameters are summarized in Table~\ref{tab:etaspectra}.
Above $\mT-\metap = 1\ \GeV$, all models result in very similar values, corresponding to an approximate \mT{}-scaling. {\it Right:} Absolute normalized spectra from input models~\cite{alcor,kaneta,rafelski,stachel} 
are compared to PHENIX 200 GeV central Au+Au collision measurements~\cite{Adler:2006bv}.
}
\label{fig:etaspec}
\end{figure}

Since the spectrum of the $\eta$ mesons can be directly compared to measured data, it also serves as a consistency check. A comparison of enhanced $\eta$ spectra of all resonance models is given on Fig.~\ref{fig:etaspec} (left), and the fitted spectrum parameters are listed in Table~\ref{tab:etaspectra}. A connection between the \etap enhancement  and the $\eta$ enhancement can be expressed as 
\begin{equation}
f_\eta \equiv \frac{N_\eta^{*}}{N_\eta} = 1 + \left( \frac{N_\etap^{*}} {N_\etap}-1 \right) \frac{N_\etap}{N_\eta}
BR(\etap\rightarrow\eta+\pi\pi)
\end{equation}
with the last term, the total \etap to $\eta$ branching ratio being approximately 65.7\%.
For models \cite{alcor,kaneta,rafelski,stachel} we have also plotted the absolutely normalized $\eta$ spectra in Fig.~\ref{fig:etaspec} (right) compared to the measured PHENIX $\eta$ spectra. 
It is obvious from the plot that the computed enhancement affects only the $\mT<1\ \mathrm{GeV}$ part of the $\eta$ spectrum.

%
%
\begin{table}[h!tbp]
\begin{tabular}{|c|c|c|c|c|c|}
\hline Resonance model & $a$ & $b$ & $c$ & $d$ & $f_\eta$ \\
\hline
\hline
No enhancement
& 14.6 & 3.38 & 0 (fixed)& 0 (fixed) & - \\
\hline
ALCOR~\cite{alcor}
& 14.6 & 3.40 & 97.0 & 17.8 & 5.25 \\
Kaneta~{\it et al.}~\cite{kaneta}
& 14.6 & 3.38 & 54.9 & 16.2 & 3.47 \\
Letessier~{\it et al.}~\cite{rafelski}
& 14.6 & 3.38 & 84.1 & 16.9 & 4.75 \\
Stachel~{\it et al.}~\cite{stachel}
& 14.5 & 3.38 & 89.2 & 17.0 & 4.97 \\
UrQMD~\cite{urqmd}
& 14.6 & 3.41 & 148 & 17.9 & 7.49 \\
\hline 
\end{tabular}
\caption{\label{tab:etaspectra}
$\eta$ enhancement ($f_\eta$) and fit parameters of the spectra for different models of resonance abundancies. The spectra are obtained using the most probable \Binv, \meps parameters, and then fitted with a double exponential function
$y = a \mathrm{e}^{-b(\mT-\metap)} + c \mathrm{e}^{-d(\mT-\metap)}$. The spectrum without enhancement is computed with the Kaneta {\it et al.}~\cite{kaneta} resonance ratios, with \meps=\metap.
}
\end{table}

Considering all errors, $6.01 \le N_\etap \le 258$ and $1.56 \le N_\eta \le 15.4$.\footnote{%
   The upper bounds on $N_\etap$ and $N_\eta$ were calculated the following way: The (\Binv, \meps) values of all above mentioned setups of input model, $\alpha$, \Tfo, \Tcond were considered along the 1-$\sigma$ contour, and then the \meps value was shifted upward with the corresponding non-model-dependent systematic error. Then the \etap (or $\eta$) spectra were plotted for these (\Binv, \meps) pairs, and the one corresponding to the maximum enhancement was considered. The lower bound was computed similarly, with the \meps values shifted downward and the spectra with the minimal enhancement taken. The maximal enhancement is given by Ref.~\cite{rafelski}, while the minimal is by Ref.~\cite{kaneta}. (Ref.~\cite{frirqmd} is not considered as FRITIOF fails to describe the STAR and the combined PHENIX and STAR data sets.)
}
Estimations using these enhancement factors indicate that the observed in-medium \etap mass drop is indeed a promising candidate~\cite{prl} to explain the dilepton excess reported by PHENIX in Ref.~\cite{Afanasiev:2007xw}.


\section{Conclusion}
The best simultaneous description of STAR and PHENIX HBT data is achieved
with an $\etap$ mass that is dramatically reduced from 958 MeV to
$340{+50\atop -60}{+280\atop -140}\pm{42}$ MeV in the medium created in central Au+Au collisions at RHIC. 
The first error here is the statistical one from the fit, the second error is from the model and parameter choices, while the third is the systematics from the centrality selection, the resolvability of $\omega$ decay products and the pseudo-rapidity cutoff.
Note that the dominant error corresponds to the selection of the model for the hadronic multiplicities.

Note that the best estimated value for the modified \etap mass does not differ significantly from the range of $412\ \MeV-715\ \MeV$, predicted by the quark model calculations for the $U_A(1)$ symmetry restoration~\cite{kapusta}. In fact, it is slightly below this range, but above the lower limit of $\sqrt{3} m_\pi$ by Weinberg~\cite{Weinberg:1975ui}. Hence the mass reduction effect seems to be already at maximum at $\sqrtsnn=200\ \GeV$ central Au+Au collisions. As a consequence, one may expect a saturation of this effect if the bombarding energy is further increased up to LHC energies of $\sqrtsnn=10\ \mathrm{TeV}$. 
In Appendix~\ref{sec:centrality} we also noted an interesting centrality dependence of the order of 9.8\%,
that suggests that the in-medium \etap mass decrease is slightly larger in more central collisions and suggests that more detailed centrality and transverse mass dependent measurements of the Edgeworth and other model-independent estimates
of the extrapolated intercept parameter $\lambda_*$ are necessary to pin down this effect, which at present is part of the above given 42 MeV systematic error.

Not only did we investigate the best value for the in-medium mass of \etap, but also the minimum mass modification 
that is required to describe the data. Based on the combined STAR and PHENIX data, and from the systematic 
investigation of various resonance multiplicities and model parameters, we conclude that in $\sqrtsnn=200\ \GeV$ central Au+Au reactions the mass of the $\etap$ meson is reduced by more than 200 MeV, at the 99.9 \% confidence 
level in the considered model class. This result was briefly summarized and published
in Ref.~\cite{prl}. A similar analysis of NA44 S+Pb data at top CERN SPS energies provided no 
evidence of an in-medium $\etap$ mass modification~\cite{vance}. For further details on the
extrapolation techniques and on  a summary of  earlier results on a correlation search for partial
$U_A(1)$ symmetry restoration we recommend the review paper~\cite{Csorgo:1999sj}.  Theoretical results and
earlier experimental searches for in-medium mass modifications of hadrons were summarized recently in Ref.~\cite{Hayano:2008vn}.

Our positive results on a significant, indirectly observed in-medium $\etap$ mass modification
should revitalize theoretical interest in other signatures of partial $\mathrm{U_A}(1)$ and chiral symmetry restoration in heavy ion reactions and also should be cross-checked against other observables like the enhancement of low-mass dileptons and photons
in the same reactions. More detailed and more precise experimental data are needed on the intercept parameter of Bose--Einstein correlations of pions at low $p_T$ at various bombarding energies, system sizes and centralities to fully understand the nature of partial $\mathrm{U_A}(1)$ symmetry restoration. Detailed analyses of other decay channels of the $\etap$ and $\eta$ mesons, e.g.\ dilepton measurements are required to confirm our observations on the onset of this effect.


\begin{acknowledgments} 
T.~Cs. is grateful to Professor R.~J.~Glauber for his kind hospitality at Harvard University and acknowledges financial support from HAESF.
We thank professor R.~J.~Glauber, M.~I.~Nagy, S.~Vainstein and Gy.~Wolf for useful discussions.
Our research was supported by OTKA grant nos.\ T49466 and NK 73143. 
\end{acknowledgments}


\appendix

\section{\label{sec:starlam}STAR Edgeworth \lambdas}
\begin{figure}[h!tbp]
\begin{center}
\includegraphics[width=.6\linewidth]{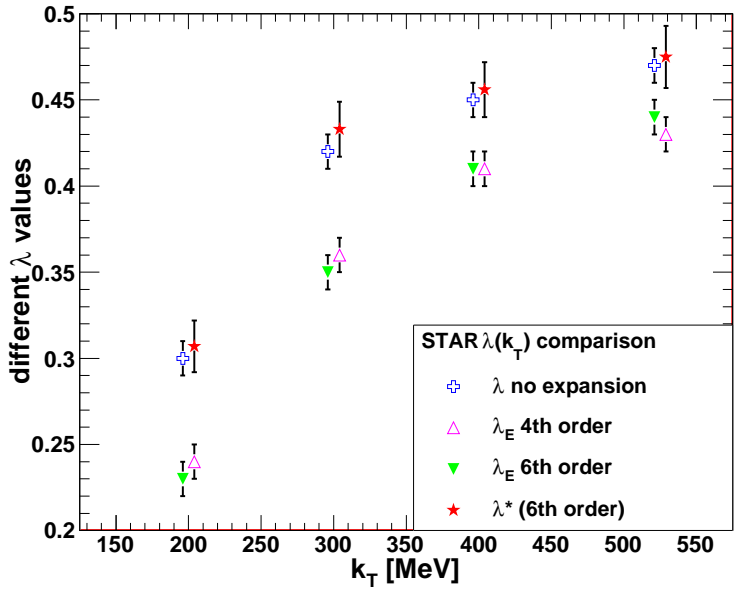}
\end{center}
\caption{{\it (Color online)}
  HBT Gaussian $\lambda$, fourth and sixth order Edgeworth $\lambda_E$ parameters from Ref.~\cite{starpub}, and the computed $\lambdas$ parameters  are shown for the four transverse momentum bins. (Note that, for the sake of visibility, the points are slightly shifted horizontally to the left or right.)
}
\label{fig:starlam}
\end{figure}

STAR used a sixth order Edgeworth expansion with the even order terms only, assuming that the source is a symmetric and analytic function. 
The STAR measurements for the $\lambda_E$ and $\kappa$ values are given in Table~\ref{tab:starlam}. Plotted values of the Gauss $\lambda$ and the fourth and sixth order $\lambda_E$, compared to our \lambdas calculations, are shown in Fig.~\ref{fig:starlam}. In this Appendix we outline the computation of  \lambdas, using the notation of STAR \cite{starpub}. Although the Hermite polynomials were defined differently from Eq.~(\ref{eq:herm}) of Ref.~\cite{starpub}, the difference cancels from the equations quoted below. The $o$, $s$, $l$ indices denote the out, side and long directions as defined at the same place.
Then the \lambdas is computed as
\begin{eqnarray}
\lambdas=&\lambda_E
\left(1+\frac{1}{8}\kappa_{o,4}-\frac{1}{48}\kappa_{o,6}\right)
\left(1+\frac{1}{8}\kappa_{s,4}-\frac{1}{48}\kappa_{l,6}\right)
\left(1+\frac{1}{8}\kappa_{l,4}-\frac{1}{48}\kappa_{s,6}\right),
\end{eqnarray}
and its error is given as
\begin{eqnarray}
\left( \Delta\lambdas \right)^2
&=&\left(\frac{\partial\lambdas}{\partial\kappa_{o,4}}\Delta\kappa_{o,4}\right)^2
+ \left(\frac{\partial\lambdas}{\partial\kappa_{o,6}}\Delta\kappa_{o,6}\right)^2
+ \left(\frac{\partial\lambdas}{\partial\kappa_{s,4}}\Delta\kappa_{s,4}\right)^2
+ \left(\frac{\partial\lambdas}{\partial\kappa_{s,6}}\Delta\kappa_{s,6}\right)^2\nonumber\\
&+&\left(\frac{\partial\lambdas}{\partial\kappa_{l,6}}\Delta\kappa_{l,4}\right)^2
+ \left(\frac{\partial\lambdas}{\partial\kappa_{l,6}}\Delta\kappa_{l,6}\right)^2
+ \left(\frac{\partial\lambdas}{\partial\lambda_E}\Delta\lambda_E\right)^2 .
\end{eqnarray}
The result of the numerical computations are summarized in Table \ref{tab:starlam}.
%
%
\begin{table}[h!tbp]
\begin{tabular}{|c|r@{}c@{}l|r@{}c@{}l|r@{}c@{}l|r@{}c@{}l|}
\hline
$\pt$ & 
\multirow{2}{*}{$150$} & \multirow{2}{*}{$-$} & \multirow{2}{*}{$250$} &
\multirow{2}{*}{$250$} & \multirow{2}{*}{$-$} & \multirow{2}{*}{$350$} &
\multirow{2}{*}{$350$} & \multirow{2}{*}{$-$} & \multirow{2}{*}{$450$} &
\multirow{2}{*}{$450$} & \multirow{2}{*}{$-$} & \multirow{2}{*}{$600$} \\ 
(MeV/c) & & & & & & & & & & & & \\
\hline\hline
 $\lambda$   & $0.30$&$\pm$&$0.01$ & $0.42$&$\pm$&$0.01$ & $0.45$&$\pm$&$0.01$ & $0.47$&$\pm$&$0.01$ \\ 
\hline
 $\lambda_E$ & $0.23$&$\pm$&$0.01$ & $0.35$&$\pm$&$0.01$ & $0.41$&$\pm$&$0.01$ & $0.44$&$\pm$&$0.01$ \\ 
 $\kappa_{o,4}$ & $0.53$&$\pm$&$0.11$ & $0.45$&$\pm$&$0.10$ & $0.20$&$\pm$&$0.11$ & $0.22$&$\pm$&$0.13$  \\
 $\kappa_{o,6}$ & $0.83$&$\pm$&$0.39$ & $0.53$&$\pm$&$0.38$ & $0.63$&$\pm$&$0.44$ & $-0.84$&$\pm$&$0.53$ \\
 $\kappa_{s,4}$ & $0.99$&$\pm$&$0.10$ & $0.79$&$\pm$&$0.10$ & $0.16$&$\pm$&$0.11$ & $-0.07$&$\pm$&$0.13$ \\
 $\kappa_{s,6}$ & $3.07$&$\pm$&$0.35$ & $3.21$&$\pm$&$0.37$ & $1.71$&$\pm$&$0.44$ & $1.80$&$\pm$&$0.51$  \\
 $\kappa_{l,4}$ & $1.32$&$\pm$&$0.07$ & $0.70$&$\pm$&$0.07$ & $0.54$&$\pm$&$0.09$ & $0.32$&$\pm$&$0.11$  \\
 $\kappa_{l,6}$ & $-1.76$&$\pm$&$0.29$ & $-2.82$&$\pm$&$0.29$ & $-2.41$&$\pm$&$0.35$ & $-2.12$&$\pm$&$0.43$ \\
\hline
 $\lambdas$ & $0.307$&$\pm$&$0.015$ & $0.433$&$\pm$&$0.015$ & $0.456$&$\pm$&$0.016$ & $0.475$&$\pm$&$0.018$ \\ 
\hline
\end{tabular}
\caption{\label{tab:starlam}
HBT Gaussian $\lambda$ parameters, $\lambda_E$ and $\kappa_{i,n}$ fit parameters of the sixth order even Edgeworth expansion for the 
5\% most central events as found by STAR \cite{starpub}, used to obtain the also shown $\lambdas$ parameters. (We kept one extra decimal here in order to avoid fitting errors later due to the coarse roundoff of the uncertainties.)
}
\end{table}

\section{\label{sec:centrality}System size, energy and centrality dependence}

\begin{figure}[h!tbp]
\begin{center}
\includegraphics[width=.6\linewidth,height=.5\linewidth]{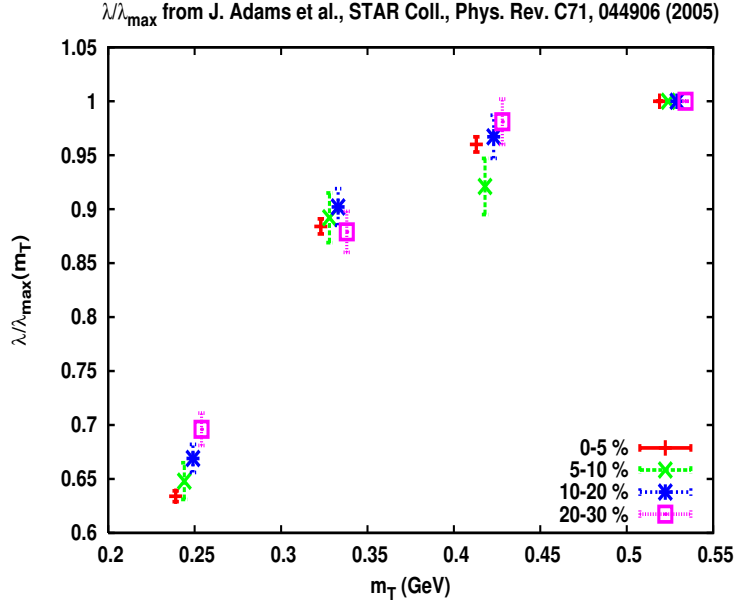}
\end{center}
\caption{{\it (Color online)}
  STAR HBT Gaussian $\lambda(\mT)/\lambda_{max}$ values for different centrality classes between 0\% and 30\% \cite{starpub}.
  The points with different centralities  were measured in the same $\mT$ intervals, but are slightly shifted to left and right on this Figure for a better visibility. Note the trend that more central data correspond to a larger hole in the lowest transverse mass bin, suggesting a slightly larger in-medium $\etap$ mass decrease in more central collisions. This effect is part of the systematic errors given in the conclusions.
}
\label{fig:starcent}
\end{figure}

The STAR Edgeworth \lambdas(\mT) data are given for the 0--5\% most central data, while PHENIX carried out the pion correlation analysis on the 0--30\% centrality range. The behavior of the \lambdas(\mT) curve is not necessarily the same for different centrality classes, resulting in a systematic error in our analysis. Fortunately, STAR provided a centrality dependent Gaussian $\lambda$ measurement, that we could use to estimate this error. 
The $\lambda(\mT)/\lambda_{max}$ values for the centrality classes between 0\% and 30\% are summarized in 
Fig.~\ref{fig:starcent}. The relative error is determined to be less than 9.8 \% given by the the difference of the first $\lambda(\mT)/\lambda_{max}$ datapoints of the 0--5\% with respect to the 20--30\% centrality classes. This is a conservative estimation of 
the  difference between 0--5\% and 0--30\% centrality data.

According to this centrality and transverse mass dependent Gaussian STAR data on the intercept parameter, 
the depth of the low-\mT dip deepens in the case of more central collisions. 
This suggests that the restoration of the \UA symmetry is more complete in case of more central collisions. 
Therefore it may be useful to measure also the more relevant Edgeworth \lambdas(\mT) data in different centrality classes too.

\begin{figure}[h!tbp]
\begin{center}
\includegraphics[width=.6\linewidth]{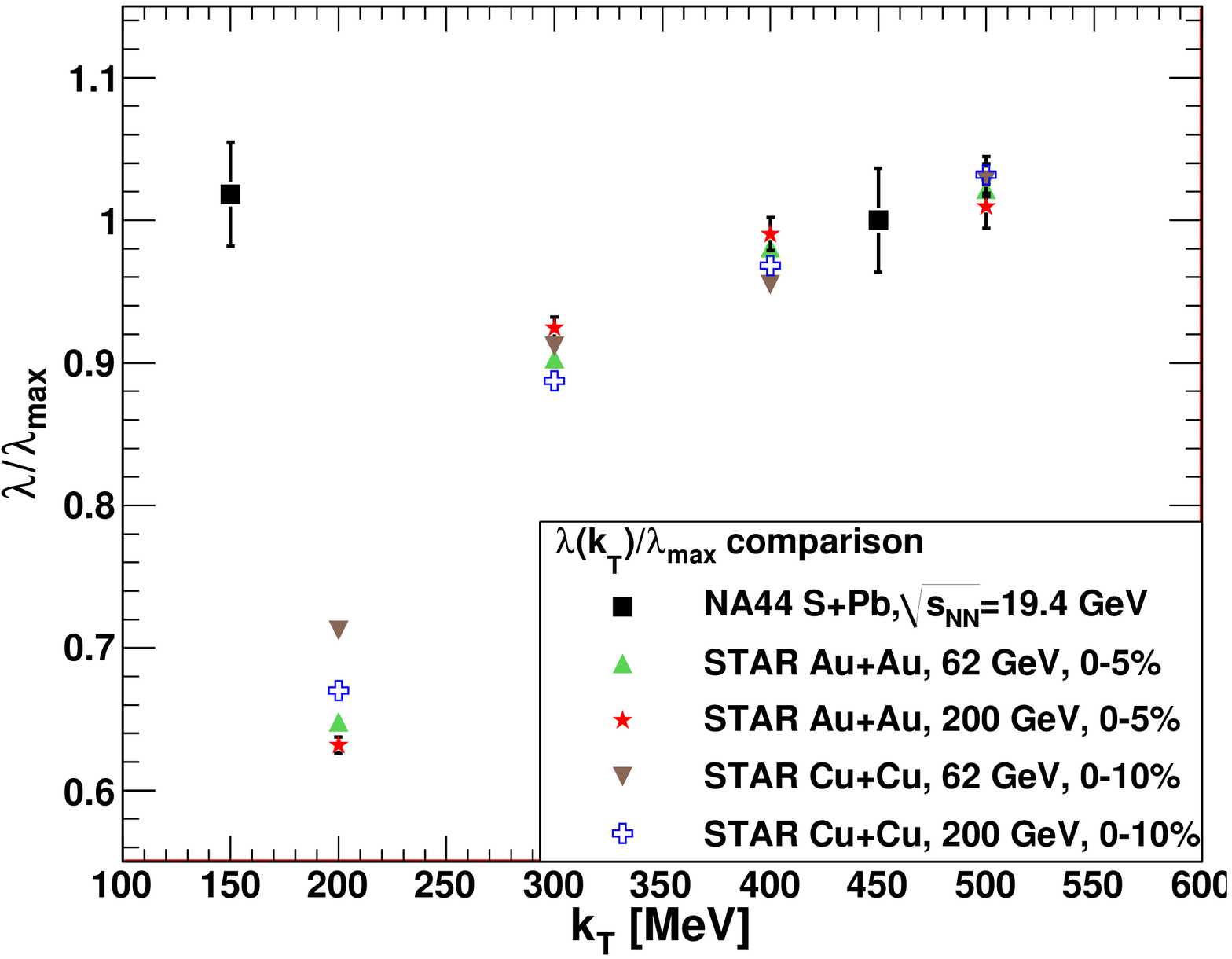}
\end{center}
\caption{{\it (Color online)}
  STAR Gaussian $\lambda(k_\mathrm{T})/\lambda_{max}$ values for RHIC collisions of different systems and energies~\cite{Abelev:2009tp}, compared to SPS NA44 measurements~\cite{Beker:1994qv}.
}
\label{fig:sysene}
\end{figure}

System size and energy dependence is shown in Fig.~\ref{fig:sysene}, where the $\lambda(k_\mathrm{T})/\lambda_{max}$ values are compared for NA44 S+Pb collisions, and STAR Au+Au and Cu+Cu collisions. The plot indicates that the mass modification effect seems to be maximal in 200 GeV Au+Au collisions, followed by 62 GeV Au+Au collisions, 200 GeV Cu+Cu and 62 GeV Cu+Cu in that order. Although in each of the STAR cases a positive signal is apparent, we also observe the lack of saturation of $\lambda(k_\mathrm{T})/\lambda_{max}$ at higher momenta, which we attribute to the changing (decreasing) amount of the radial flow with decreasing energy and system size. Such a dependence on the amount of the radial flow has been first pointed out in Ref.~\cite{vance}.

\section{\label{sec:models}Models}
%
\begin{table}
\begin{tabular}{|ll|c|c|c|c|c|c|}
\hline
\multicolumn{2}{|c|}{\multirow{3}{*}{Particles}} & \multicolumn{6}{c|}{Fractions from model} \\
\multicolumn{2}{|c|}{} 	    		    & ALCOR  & FRITIOF & Kaneta & Letessier & Stachel & UrQMD \\ 
 & & \cite{alcor} & \cite{frirqmd} & {\it et al.}~\cite{kaneta} & {\it et al.}~\cite{rafelski} & {\it et al.}~\cite{stachel} & \cite{urqmd} \\
\hline\hline
primordial 		    & $\pip$ 	    & 0.4910 & 0.2095 & 0.7396 & 0.3059 & 0.3333 & 0.2395 \\ 
\hline
\multirow{5}{*}{short lived}& $\rho$ 	    & 0.1100 & 0.3058 & 0.0651 & 0.0259 & 0.0370 & 0.0045 \\
			    & $\Delta$ 	    & 	     & 0.0846 & 0.0088 & 0.0080 &	 & 0.0069 \\
			    & $\mathrm{K}^*$& 	     & 0.1351 & 0.0124 & 0.0143 &	 & 0.0014 \\
			    & $\Sigma$      &	     & 0.0153 & 0.0040 & 0.0017 &  	 & 0.0066 \\
			    & $\Sigma^*$    &	     & 0.0098 & 0.0054 &        &  	 & 0.0016 \\ 
\hline
\multirow{4}{*}{long lived} & $\omega$ 	    & 0.1104 & 0.1023 & 0.0209 & 0.0233 & 0.0296 & 0.0073 \\
			    & $\eta$	    & 0.0453 & 0.0516 & 0.0602 & 0.0383 & 0.0360 & 0.0233 \\
			    & $\etap$	    & 0.0067 & 0.0577 & 0.0089 & 0.0032 & 0.0031 & 0.0050 \\
			    & $\ks$ 	    & 0.0717 & 0.0283 & 0.0746 & 0.0601 & 0.0513 & 0.0287 \\
\hline
\end{tabular}
\caption{\label{tab:resfrac}Resonance ratios from different models.}
\end{table}
In this section we provide a brief description---both general and analysis specific---of the six different models used to simulate the number of each important resonance that decays into pions, as well as the number of primordial pions. The fractions of pions from different sources are compared to the total number of particles in Table \ref{tab:resfrac}. These values are fed into our simulations as an input.
Note that not all core resonances are listed. With the assumption of a fixed \uT, the $\lambdas(\mT)$ is solely determined by the number of the halo resonances ($\omega$, $\eta$, $\etap$, $\ks$), \textit{and} by the total number of pions. The primordial pions can replace those pions that come from fast decays without any effect on $\lambdas(\mT)$. Generally, if a model contains a large number of exotic short lived resonances, it predicts less primordial pions.

\textbf{ALCOR}~\cite{alcor}:
Hadron multiplicities---especially for strange particles---are calculated in the framework of this algebraic coalescence rehadronization model, which counts for redistribution of quarks into hadrons for relativistic heavy-ion collisions.
Resonance ratios were taken from recent calculations for the RHIC collisions~\cite{Levai:2008me}. Since the mass eigenstates of the strange-antistrange mesons are not directly pinned down, the Kaneta-Xu~\cite{kaneta} predictions were used to fix the $N_{\etap}/N_{\eta}$ ratio for the ALCOR model.

\textbf{FRITIOF}~\cite{frirqmd}
is a Monte Carlo program that implements the Lund string model for hadron-hadron, hadron-nucleus and nucleus-nucleus collisions. Resonance ratios were computed with the FRITIOF model by simulating of 1000 events using RHIC central Au+Au $\sqrtsnn=200$ GeV parameters. FRITIOF was excluded from examination when drawing the conclusions of this analysis, since it was unable to describe the STAR \lamfrac data set, nor the combined STAR + PHENIX data set, to a $CL>0.1\%$ at any setup of the model parameters.

\textbf{Kaneta-Xu}~\cite{kaneta}:
In the case of this model, Eq.~(1). of \cite{kaneta} was used the following way:
\begin{equation}\label{eq:kaneta}
\rho_i=\frac{2J_i+1}{2\pi^2}T_{ch}m_i^2K_2\left(\frac{m_i}{T_{ch}}\right)
\end{equation}
where $m_{i}$ is the mass, $J_{i}$ is the spin of the particle. $K_2(x)$ denotes the second order modified Bessel function. For the sake of simplicity the temperature of the chemical equilibrium $T_{ch}=160\ \MeV$, the strangeness saturation factor $\gamma^{s}=1$, the quark potentials $\mu_q=10\ \MeV$ and $\mu_q=0$ were considered, all in consistency with the published PHENIX and STAR measurements in $\sqrtsnn=200\ \GeV$ Au+Au collisions~\cite{Adcox:2004mh}. (The relative minuteness of $\mu_q$ guarantees that the charge factor is negligible for even those rare particles, where it is not exactly 0). 

\textbf{Letessier-Rafelski}~\cite{rafelski}:
The model of these two authors studied soft hadron production in relativistic heavy ion collisions in a wide range of reaction energy, $4.8\ \GeV < \sqrtsnn < 200\ \GeV$, and made predictions about yields of particles using the statistical hadronization model. Particle yields of the central events were taken from Table~4 of \cite{rafelski}.

\textbf{Stachel \it{et\ al.}}~\cite{stachel}:
This statistical ``fireball'' model treats the system as a grand canonical ensemble with the temperature and the baryon chemical potential as free parameters. Particle yields of the central events were taken from Sec.~2.2. Table~1 of \cite{stachel}.

\textbf{UrQMD}~\cite{urqmd}:
The ultra-relativistic quantum molecular dynamics model is a microscopic model used to simulate (ultra)relativistic heavy ion collisons in the energy range from Bevalac and Heavy Ion Synchrotron (SIS) up to Alternating Gradient Synchrotron (AGS), SPS and RHIC. 
Resonance ratios were computed with the UrQMD model by simulating 1000 events using RHIC central Au+Au $\sqrtsnn=200$ GeV  parameters.

\end{document}